\documentclass[a4paper,12pt,dvipsnames]{article}
\pdfoutput=1

\usepackage{aas_macros}

\usepackage{jheppub}

\usepackage[utf8]{inputenc}
\usepackage{graphicx}
\usepackage{amsmath}
\usepackage{amssymb}
\usepackage{bm}
\usepackage{bbold}
\usepackage{paralist,array}
\usepackage{units}
\graphicspath{{./Plots/}}
\usepackage{slashed}
\usepackage{todonotes}
\usepackage[parfill]{parskip}
\usepackage{longtable}
\usepackage[caption=false]{subfig}
\usepackage{isotope}
\usepackage{multirow}

\graphicspath{{plots/}}

\title{An allowed window for heavy neutral leptons below the kaon mass}
\author[a,b]{Kyrylo Bondarenko,}
\author[c]{Alexey Boyarsky,}
\author[b]{Juraj Klaric,}
\author[c,e]{Oleksii Mikulenko,}
\author[d]{Oleg Ruchayskiy,}
\author[d]{Vsevolod Syvolap,}
\author[b]{and Inar Timiryasov}
\affiliation[a]{Theoretical Physics Department, CERN, 1 Esplanade des Particules, Geneva 23, CH-1211, Switzerland}
\affiliation[b]{Institute of Physics, Laboratory for Particle Physics and Cosmology,\\
\'{E}cole Polytechnique F\'{e}d\'{e}rale de Lausanne, CH-1015 Lausanne, Switzerland}
\affiliation[c]{Intituut-Lorentz, Leiden University, Niels Bohrweg 2, 2333 CA Leiden, The Netherlands}
\affiliation[d]{Niels Bohr Institute, University of Copenhagen, Blegdamsvej 17, DK-2010, Copenhagen, Denmark}
\affiliation[e]{Department of Physics, Taras Shevchenko National University of Kyiv, 64 Volodymyrs’ka str., Kyiv 01601, Ukraine}

\emailAdd{kyrylo.bondarenko@cern.ch}
\emailAdd{boyarsky@lorentz.leidenuniv.nl}
\emailAdd{juraj.klaric@epfl.ch}
\emailAdd{mikulenko@lorentz.leidenuniv.nl}
\emailAdd{oleg.ruchayskiy@nbi.ku.dk}
\emailAdd{vsevolod.syvolap@nbi.ku.dk}
\emailAdd{inar.timiryasov@epfl.ch}

\begin{document}

\abstract{
	The extension of the Standard Model with two gauge-singlet Majorana fermions can simultaneously explain two beyond-the-Standard-model phenomena:  neutrino masses and oscillations, as well as the origin of the matter-antimatter asymmetry in the Universe.
	The parameters of such a model are constrained by the neutrino oscillation data, direct accelerator searches, big bang nucleosynthesis, and requirement of successful baryogenesis. 
	We show that their combination still leaves an allowed region in the parameter space below the kaon mass.
	This region can be probed by the further searches of NA62, DUNE, or SHiP experiments.
}

\maketitle

\section{Introduction}
The observed light neutrino masses and the baryon asymmetry of the Universe (BAU) remain some of the biggest hints pointing at physics beyond the Standard Model (SM).
One of the simplest answers to both questions could be the existence of gauge-singlet Majorana fermions --- also known as right-handed neutrinos, sterile neutrinos, or heavy neutral leptons (HNLs).
HNLs can provide neutrino masses 
via the seesaw mechanism~\cite{Minkowski:1977sc,Yanagida:1979as,GellMann:1980vs,Mohapatra:1979ia,Schechter:1980gr,Schechter:1981cv}.
Right-handed neutrinos can also be responsible for the generation of the BAU through the process known as \emph{leptogenesis}, see, e.g.~\cite{Davidson:2008bu,Canetti:2012zc,Bodeker:2020ghk} for reviews and~\cite{Klaric:2020lov,Klaric:2021cpi} for a recent update.
The neutrino oscillation data requires at least two right-handed neutrinos. It turns out that the same HNLs with masses in MeV--GeV range can successfully generate the BAU~\cite{Asaka:2005pn,Canetti:2010aw,Klaric:2020lov}.
\footnote{This model can be viewed as a part of the $\nu$MSM, where the third singlet fermion plays a role of dark matter candidate~\cite{Asaka:2005an,Asaka:2005pn}. Dark Matter in the $\nu$MSM can be produced resonantly~\cite{Shi:1998km,Shaposhnikov:2008pf,Laine:2008pg,Canetti:2012kh,Ghiglieri:2020ulj}, which requires large lepton asymmetry (see, e.g.\ the recent work~\cite{Ghiglieri:2020ulj} and references therein). Alternatively, it can be produced during  preheating~\cite{Bezrukov:2011sz,Shaposhnikov:2020aen}.}

As a result of the seesaw mechanism, the SM neutrinos (flavor eigenstates) mix with the light ($\nu_i$) and heavy ($N_I$) mass eigenstates:
\begin{equation}
	\nu_{L \alpha}=V_{\alpha i}^\text{PMNS} \nu_{i}+\theta_{\alpha I} N_{I}^{c},
	\label{mixing}
\end{equation}
where $V_{\alpha i}^\text{PMNS}$ is the PMNS matrix (see, e.g.~\cite{Zyla:2020zbs}) and the matrix $\theta_{\alpha I}$ characterizes the mixing  between the HNLs and flavor states.
These mixings are subject to a number of constraints:

\begin{itemize}

\item The ratios of the mixing angles $\theta_{\alpha I}$ are constrained by the neutrino oscillation data~(see e.g.~\cite{Ruchayskiy:2011aa,Asaka:2011pb,Drewes:2016jae,Caputo:2017pit}).
Somewhat surprisingly, the values  $\theta_{\alpha I}$ themselves are not bounded from above by the neutrino oscillation data, provided that certain cancellations between these elements ensure the smallness of neutrino masses~\cite{Wyler:1982dd,Mohapatra:1986bd,Branco:1988ex,GonzalezGarcia:1988rw,Shaposhnikov:2006nn,Kersten:2007vk,Abada:2007ux,Gavela:2009cd}.

\item If HNLs are sufficiently light (below the electroweak scale), their existence can be probed
directly~\cite{Shrock:1980vy,Shrock:1980ct,Shrock:1981wq,Shrock:1981cq,Shrock:1982sc,Atre:2009rg,Drewes:2013gca,Gronau:1984ct,Cvetic:2013eza,Cvetic:2014nla,Cvetic:2015ura,Cvetic:2015naa,Cvetic:2018elt,Cvetic:2019rms,Deppisch:2015qwa,Zamora-Saa:2016ito,Caputo:2016ojx,Bolton:2019pcu,Das:2017zjc,Dib:2019tuj,Antusch:2017hhu,Bryman:2019bjg}.  A partial list of the searches at the existing experiments is~\cite{Liventsev:2013zz,Artamonov:2014urb,Aaij:2014aba,Khachatryan:2015gha,Aad:2015xaa,CortinaGil:2017mqf,Izmaylov:2017lkv,Sirunyan:2018mtv,Abe:2019kgx,NA62:2020mcv,NA62mupreliminary,Aad:2019kiz,Tastet:2020tzh}. HNL searches are an important part of the physics program of many proposed experiments, see, e.g.~\cite{Mermod:2017ceo,Gligorov:2017nwh,Curtin:2018mvb,Drewes:2018gkc,Zamora-Saa:2019naq,Boiarska:2019jcw,Beacham:2019nyx,Ballett:2019bgd,SHiP:2018xqw,Hirsch:2020klk}.
In the absence of  positive results, one establishes upper limits on the mixing angles.

\item The HNLs can be copiously produced in the early Universe. 
Their subsequent decays can affect light element abundances and establish an upper bound on the lifetime of HNLs, i.e.\ a lower bound on their mixing angles~\cite{Dolgov:2000pj,Dolgov:2000jw,Dolgov:2003sg,Fuller:2011qy,Ruchayskiy:2012si,Hernandez:2013lza,Hernandez:2014fha,Vincent:2014rja,Gelmini:2019wfp,Kirilova:2019dlk,Gelmini:2020ekg,Sabti:2020yrt,Boyarsky:2020dzc, Boyarsky:2021yoh}.
\item Finally, as we already mentioned, HNLs can be responsible for the generation of the BAU.
Leptogenesis in this model has attracted significant interest of theoretical community and several groups have performed studies of the parameter space~\cite{Canetti:2012vf,Canetti:2012kh,Shuve:2014zua,Abada:2015rta,Hernandez:2015wna,Drewes:2016gmt,Drewes:2016jae,Hernandez:2016kel,Hambye:2017elz,Abada:2017ieq,Antusch:2017pkq,Ghiglieri:2017csp,Eijima:2018qke,Ghiglieri:2018wbs,Klaric:2020lov,Eijima:2020shs,Klaric:2021cpi}. The requirement of successful leptogenesis limits the values of the mixings $\theta_{\alpha I}$ from above and from below.
\end{itemize}

Experimental studies usually report their results in terms of a single HNL mixing with a single flavor. While convenient for comparison between studies, such a model of HNL does not solve \textit{per se} any of the BSM problem outlined above -- neutrino masses require at least two HNLs, these HNLs should be mixed with several flavors to explain oscillations; low-scale leptogenesis also requires at least two HNLs with sufficient mass degeneracy to enhance the production of the baryon asymmetry.
The simplest HNL model capable of incorporating the above BSM phenomena, is the model with two HNLs, having approximately equal masses ($M_{N_1} \approx M_{N_2} = M_N$).
It turns out that the parameter space of such a model looks quite different from a parameter space of a toy-model with a single HNL. 
To correctly combine different constraints  one has to reanalyze different experimental data. 
Our work is devoted to such a reanalysis for the model with two degenerate  HNLs.

Similar reinterpretation works have been performed in the past.
In Ref.~\cite{Drewes:2015iva} collider searches were combined with the constraints from the seesaw mechanism for three heavy neutrinos, as well as the limits on HNL lifetime from BBN (see also~\cite{Chrzaszcz:2019inj} for a recent analysis using the GAMBIT~\cite{Athron:2017ard} framework).
The minimal model with two HNLs is more restrictive and leads to stronger bounds on the properties of HNLs, even more so when combined with the condition of successful leptogenesis as was done in Ref.~\cite{Drewes:2016jae}. 
Ref.~\cite{Hernandez:2016kel} also discussed constraints on the parameter space of the two HNL  model, taking into account leptogenesis as well as potential future searches for the HNLs and for the neutrinoless double beta decay signal.

In this work we revise and improve upon the existing studies in a number of ways:
\begin{enumerate}
    \item There has been a significant improvement of the Big Bang Nucleosynthesis constraints,
     based on the meson driven conversion effect~\cite{Boyarsky:2020dzc} (as compared to $\tau<\unit[0.1]{s}$ bound in Ref.~\cite{Drewes:2015iva, Drewes:2016jae, Chrzaszcz:2019inj}). This drastically affects our study  rendering out a significant portion of the parameter space.
    \item We include the state-of-the-art leptogenesis calculations that incorporate important processes, neglected in previous studies~(see e.g.\ \cite{Eijima:2017anv, Ghiglieri:2017gjz}).
    \item The allowed mixing angles are significantly affected by the improved bounds on the CP violating phase $\delta_{\text{CP}}$~\cite{Esteban:2020cvm}, which was not constrained in the previous studies.
    \item We include the latest results of the direct searches from the NA62 experiment~\cite{NA62:2020mcv}.
\end{enumerate}
Given the impact of the works from recent years, it is important to see how the combination of these constraints is affected, and to reevaluate the remaining parameter space,
especially given the attention of the community towards feebly interacting particles (FIPs)~\cite{Agrawal:2021dbo}. 
In this work we present the most up-to-date bounds on the properties of HNLs in the minimal model.

The paper is organized as follows: in Section~\ref{sec:oscillations} and~\ref{sec:accelerators} we discuss constraints on the model from the neutrino oscillation data and accelerator searches. In Section~\ref{sec:constraints_from_big_bang_nucleosynthesis} we discuss constraints from BBN, taking into account the effects of mesons produced from the HNL decays, while in Section~\ref{sec:constraints_from_leptogenesis} we present the parameter region where BAU could be successful. We combine these limits in Section~\ref{sec:combined-limits} and give a discussion in Section~\ref{sec:discussion_and_outlook}.

\section{Constraints from  neutrino oscillations}
\label{sec:oscillations}
The experimentally observed neutrino oscillations cannot be explained within the Standard Model of particles physics where neutrinos are massless and the flavor lepton number is conserved. 
One of the possible ways to solve this problem is to add two right-handed neutrinos to the model.
In addition to the Dirac masses $m_D = v F$, which couple the left-handed and right-handed neutrinos, the right-handed neutrinos, being gauge-singlet states, can also have Majorana masses $M_M$, unrelated to the SM Higgs field.
To find the physical states we need to diagonalise the full mass matrix of the left- and right-handed neutrinos
\begin{align}
	\mathcal{L} \supset
	\frac12
	\begin{pmatrix}
		\overline{\nu_L} & \overline{\nu_R^c}
	\end{pmatrix}
	\begin{pmatrix}
		0 & m_D\\
		m_D^T & M_M
	\end{pmatrix}
	\begin{pmatrix}
		\nu_L^c \\ \nu_R
	\end{pmatrix}\,.
\end{align}
If the Dirac masses are small compared to the Majorana masses, we can block-diagonalise this matrix to find two sets of masses,
\begin{align}
	m_\nu \simeq m_D M_M^{-1} m_D^T\quad \text{and}\quad
	M_M \simeq \begin{pmatrix}
	    M_1 & 0 \\ 0 & M_2
	\end{pmatrix}\,.
	\label{eq:seesaw}
\end{align}
This is the famous \emph{seesaw} formula~\cite{Minkowski:1977sc,Yanagida:1979as,GellMann:1980vs,Mohapatra:1979ia,Schechter:1980gr,Schechter:1981cv}, in which the smallness of the light neutrino masses $m_\nu$ is explained by the parametrically small ratio of the Dirac and Majorana masses $m_D/M_{1,2} \ll 1$.
Another consequence of the seesaw mechanism is that the heavy states (henceforth \emph{heavy neutral leptons} -- HNLs) are mixtures of the left-handed and right-handed neutrinos, and can interact with the rest of the SM -- in particular with the $W$ and $Z$ bosons.
The strength of this interaction is given by the \emph{mixing angle}:
\begin{align}
	\theta \simeq m_D M_M^{-1} && m_\nu = \theta M_M \theta^T\,.
	\label{def:mixingAngle}
\end{align}

The modulus squared of the mixing angle quantifies how suppressed the HNL interactions are compared to the interactions of the light neutrinos.
It is often useful to introduce the quantities
\begin{subequations}
	\begin{align}
		\label{def:U2I}
		&U_{\alpha I}^2 \equiv |\theta_{\alpha I}|^2\,, && U_I^2 \equiv \sum_\alpha U_{\alpha I}^2 \,,\\
		\label{def:U2}
		&U^2_\alpha \equiv \sum_I U^2_{\alpha I}\,, && U^2 \equiv \sum_{\alpha I} U_{\alpha I}^2\,,
	\end{align}
\end{subequations}
which quantify the overall suppression of the HNL interactions. If the HNLs are degenerate in mass, it is also useful to consider the sum over the HNL flavors~\eqref{def:U2}.

It is important to note that the size of the observed neutrino masses $m_\nu$ does not constrain the mixing angles $\theta$, nor the Majorana mass $M_M$, but only their combination from Eq.~\eqref{def:mixingAngle}.
This suggests that the seesaw mechanism does not imply a mass scale for the heavy neutrinos.
Nonetheless, using the seesaw relation~\eqref{eq:seesaw}, we can connect the HNL mixing angles $\theta$ to the known neutrino oscillation data through the Casas-Ibarra parametrization~\cite{Casas:2001sr}:
\begin{align}
	\theta = i V^{\text{PMNS}} \left(m_\nu^\mathrm{diag}\right)^{1/2} R \left(M_M^\mathrm{diag}\right)^{-1/2}\,,
	\label{CI:Yukawa}
\end{align}
where $V^{\text{PMNS}}$ is the PMNS matrix, $m_\nu^\mathrm{diag}$ is the light neutrino mass matrix with $m_1=0$ for normal hierarchy (NH),\footnote{In the model with two HNLs which we consider here the lightest active neutrino is massless at tree level~\cite{Davidson:2006tg} and therefore we use the term \textit{hierarchy} rather than ordering.} and $m_3=0$ for inverted hierarchy (IH).
The complex matrix $R$ satisfies the relation $R^T R = \mathbb{1}_{2\times2}$,
and depends on the neutrino mass hierarchy
\begin{align}
	R^{\rm NH}=
	\begin{pmatrix}
		0 && 0\\
		\cos \omega && \sin \omega \\
		-\xi \sin \omega && \xi \cos \omega
	\end{pmatrix}\,, &&
	R^{\rm IH}=
	\begin{pmatrix}
		\cos \omega && \sin \omega \\
		-\xi \sin \omega && \xi \cos \omega \\
		0 && 0
	\end{pmatrix}
	\,,
\end{align}
where $\omega$ is a complex-valued angle, and $\xi = \pm 1$. The symmetry $\xi \leftrightarrow -\xi$, $N_1 \leftrightarrow N_2$, $\omega \leftrightarrow \omega +\frac{\pi}{2}$ allows to consider only the $\xi = +1$ case.

While this parametrization cannot provide a limit on the individual mixing angles $U^2_{\alpha I}$~\cite{Drewes:2019mhg},  in the degenerate mass limit it gives a lower bound on the summed mixing angles, see e.g.~\cite{Ruchayskiy:2011aa,Eijima:2018qke}.
If the HNL mixing angle is large $U^2_\alpha \gg m_\nu/M_N$\footnote{
If $|m_D|^2/M_N^2 \approx U^2 = \mathcal{O}(1)$, the seesaw expansion breaks down, and one should go beyond Casas-Ibarra parametrization~\cite{Donini:2012tt}.
Given the strong experimental constraints on $U^2$, we can safely neglect such a correction in the present analysis.},  two HNLs form a quasi-Dirac pair with mixing angles that are approximately equal up to a phase $\theta_{\alpha 2} \approx i \theta_{\alpha 1}$,
and the expression for $U^2_\alpha$ is given by~\cite{Shaposhnikov:2008pf,Asaka:2011wq,Ruchayskiy:2011aa,Hernandez:2016kel,Drewes:2016jae}:
\begin{align}
	\label{eq:mixing pattern casasibarra}
	U_{\alpha}^2 = |U_{\alpha 1}|^2+|U_{\alpha 2}|^2 & \approx \frac{e^{2|\text{Im } \omega|}}{2 M_N} 
	\Bigl(m_2 |V^{\text{PMNS}}_{\alpha 2}|^2 + m_3 |V^{\text{PMNS}}_{\alpha 3}|^2 \nonumber \\
	&-\text{sgn(Im}\,\omega) \cdot 2 \sqrt{m_2 \, m_3} \text{Im} [V^{\text{PMNS}}_{\alpha 2}(V^{\text{PMNS}}_{\alpha 3})^*] \Bigr ),
\end{align}
for the NH while for the IH case the r.h.s. of the corresponding equations are obtained by replacing $2\rightarrow1$ and $3\rightarrow2$.
When we normalize the flavored mixing angles $U^2_\alpha$ to the total mixing angle $U^2$,
the dependence on the unknown HNL parameters drops out, and the ratios $U^2_\alpha/U^2$
depend only on the PMNS parameters~\cite{Drewes:2016jae,Caputo:2017pit}.
In what follows we will encounter these ratios very often, so we introduce
\begin{equation}
    x_\alpha \equiv U^2_\alpha/U^2, \qquad x_e+x_\mu+x_\tau = 1.
    \label{x_def}
\end{equation}
The Majorana phases entering the PMNS matrix, $\alpha_{21}$ and $\alpha_{31}$ (in PDG conventions), which also affect 
the ratios~\eqref{x_def}, cannot be determined in oscillation experiments, but could instead be measured indirectly through neutrinoless double beta decay experiments in the near future (see, e.g.~\cite{Bezrukov:2005mx,Drewes:2016lqo,Asaka:2016zib,Hernandez:2016kel}).
In the limit of two HNLs, there is effectively only one Majorana phase which we denote $\eta$. The phase $\eta$ is equal to  $(\alpha_{21} - \alpha_{31})/2$ in the case of normal hierarchy and $\alpha_{21}/2$ in the case of inverted hierarchy.

\begin{figure}[h!]
\label{fig:neutr}
	\centering
	\includegraphics[width = 0.49\textwidth]{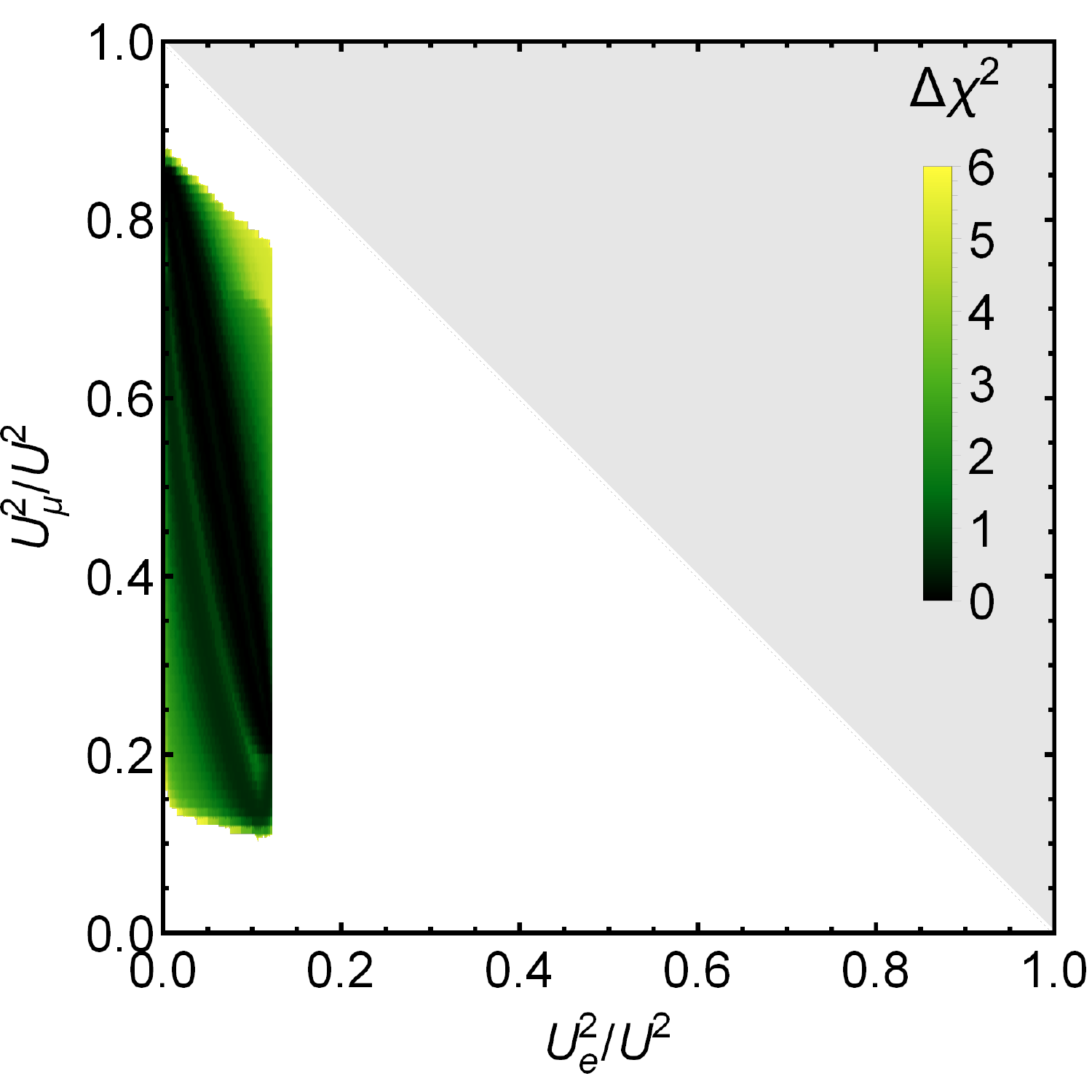}~
	\includegraphics[width = 0.49\textwidth]{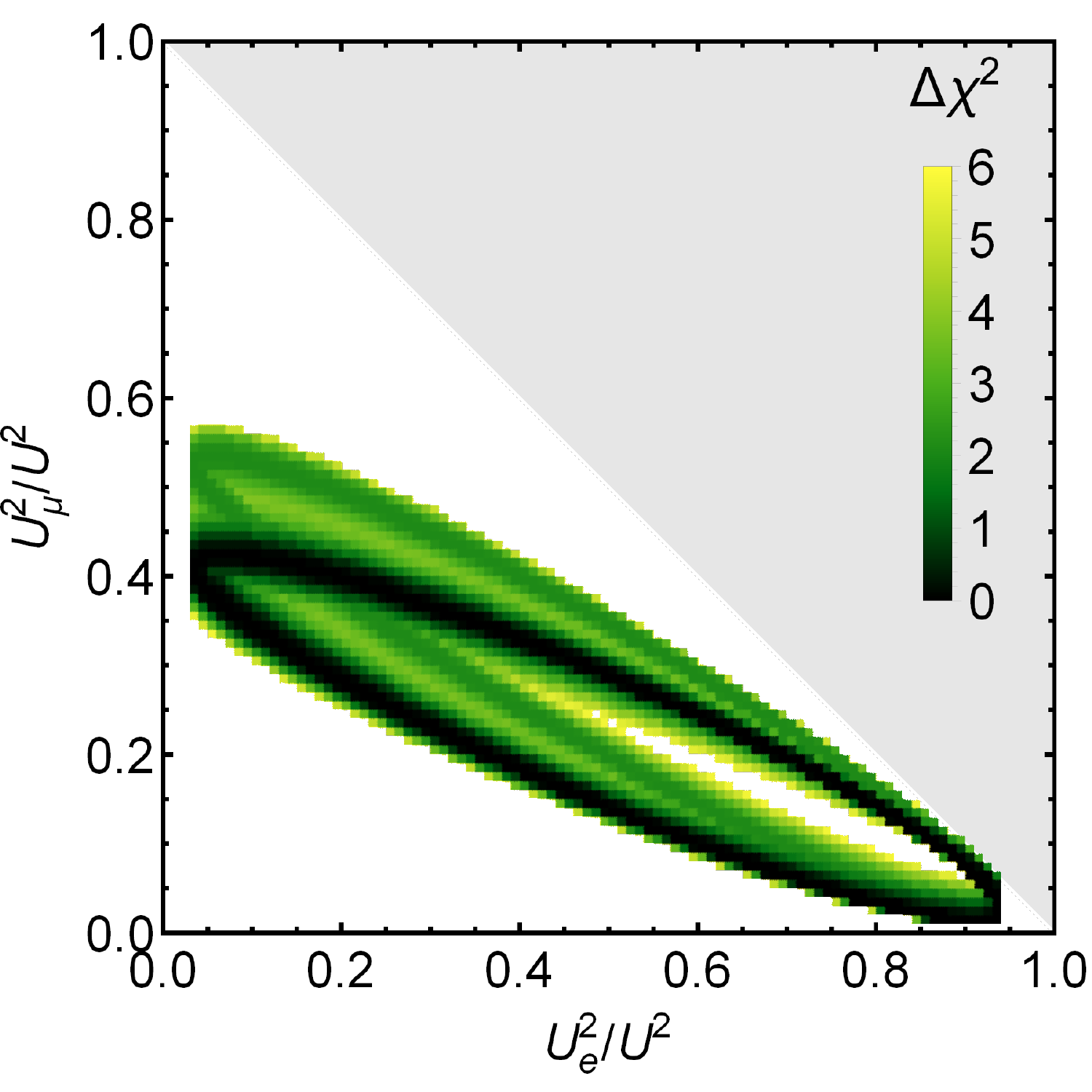}
	\caption{95\% bounds for $x_\alpha = U^2_\alpha/U^2$ for normal hierarchy (left) and inverted hierarchy (right) in the $e^{2\,\text{Im }\omega}\gg 1$ limit. The Majorana phase takes values $\eta \in [0, 2\pi)$, while $\Delta \chi^2$ is taken for the measured values of the PMNS angles $\theta_{23}$ and $\delta$, that affect the region most strongly. Gray area corresponds to the forbidden region of the parameter space $x_e + x_\mu > 1$, see Eq.~\eqref{x_def}.}
	\label{fig:oscillation-bounds}
\end{figure}

To determine the allowed mixing patterns, we take the latest neutrino oscillation parameters from nuFIT 5.0~\cite{Esteban:2020cvm} (without Super-Kamiokande atmospheric data).
We perform a numerical scan for the different values of the Dirac phase of the PMNS matrix $\delta_{\text{CP}}$, angle $\theta_{23}$ and $\eta$, as for the remaining parameters the experimental uncertainty is sufficiently smaller and their variation only slightly change the allowed parameter space.
For each point in the $x_e$--$x_\mu$ plane we find the smallest possible $\Delta \chi^2(\delta_{\text{CP}}, \theta_{23})$ and take only the points with $\Delta \chi^2<6$ which correspond to the $95\%$ region for 2 degrees of freedom.

The result of this procedure is shown in Fig.~\ref{fig:oscillation-bounds}. We see that for normal hierarchy $x_e$ can reach small values, while for inverted hierarchy all three $x_{\alpha}$ can be small. As we will see later, the results for the minimal allowed HNL mass depend on these small numbers, so one needs to determine them with high accuracy.
Therefore, we analyzed minimal values of all $x_{\alpha}$, using the two-dimensional $\Delta\chi^2$ projection from nuFIT data for the two most relevant parameters for each case. The minimal values within 2$\sigma$ bounds are given in Table~\ref{tab:minx}, c.f.~\cite{Ruchayskiy:2011aa}.

\begin{table}[t!]
    \centering
    \begin{tabular}{|c|c|c|c|c|}
    \hline
        & \multicolumn{2}{c|}{NH}   & \multicolumn{2}{c|}{IH}  \\
        \hline
        & rel. param. & min value & rel. param. & min value \\
        \hline
         $x_e$ &$\theta_{12}, \theta_{13}$ & 0.0034 &  $\theta_{12}, \Delta m_{\text{sol}}$ & $0.026$\\
         \hline
         $x_\mu$ &$\theta_{23}, \delta_{\text{CP}}$ & 0.11 &  $\theta_{12}, \delta_{\text{CP}}$ & $3.2 \cdot 10^{-4}$\\
        \hline 
         $x_\tau$ &$\theta_{23}, \delta_{\text{CP}}$ & 0.11 &  $\theta_{12}, \delta_{\text{CP}}$ & 0.0011\\ 
         \hline
    \end{tabular}
    \caption{Minimal values of $x_\alpha = U^2_\alpha/U^2$ allowed by neutrino oscillation data for both normal (NH) and inverted (IH) hierarchies. The column ``rel. param.'' shows the most relevant neutrino oscillation parameters that change the minimal $x_\alpha$ values.
    }
    \label{tab:minx}
\end{table}

\section{Constraints from accelerator experiments}
\label{sec:accelerators}

There exist two types of accelerator experiments capable of searching for MeV-GeV mass HNLs. The first type is  \emph{missing energy experiments} searching for decays $\pi/K\rightarrow e/\mu +\text{(invisible) }$.
The probability of these decays depends solely on $U^2_{e/\mu}$, directly probing mixing angles independently on the mixing pattern. The bounds obtained in this type of experiments are generally stronger than for other types of experiments, however, they can only constraint HNLs with mass lower than kaon mass. In addition, they are not sensitive to combinations $U_\alpha U_\beta$ ($\alpha\neq\beta$) and cannot constrain $U^2_\tau$ (because of large tau-lepton mass $m_\tau>m_K$). We use explicit bounds from: PIENU~\cite{Aguilar-Arevalo:2017vlf}, TRIUMPH~\cite{Britton:1992xv} ($\pi\rightarrow e$), KEK~\cite{Yamazaki:1984sj}, NA62~\cite{NA62:2020mcv,NA62mupreliminary} ($K\rightarrow e/\mu$), 
E949~\cite{Artamonov:2014urb} ($K\rightarrow \mu$). For NA62 $K\rightarrow \mu$ decay only 30\% of the current data has been processed~\cite{NA62mupreliminary}.

\begin{figure}[h!]
	\centering
	\includegraphics[width = 0.47\textwidth]{"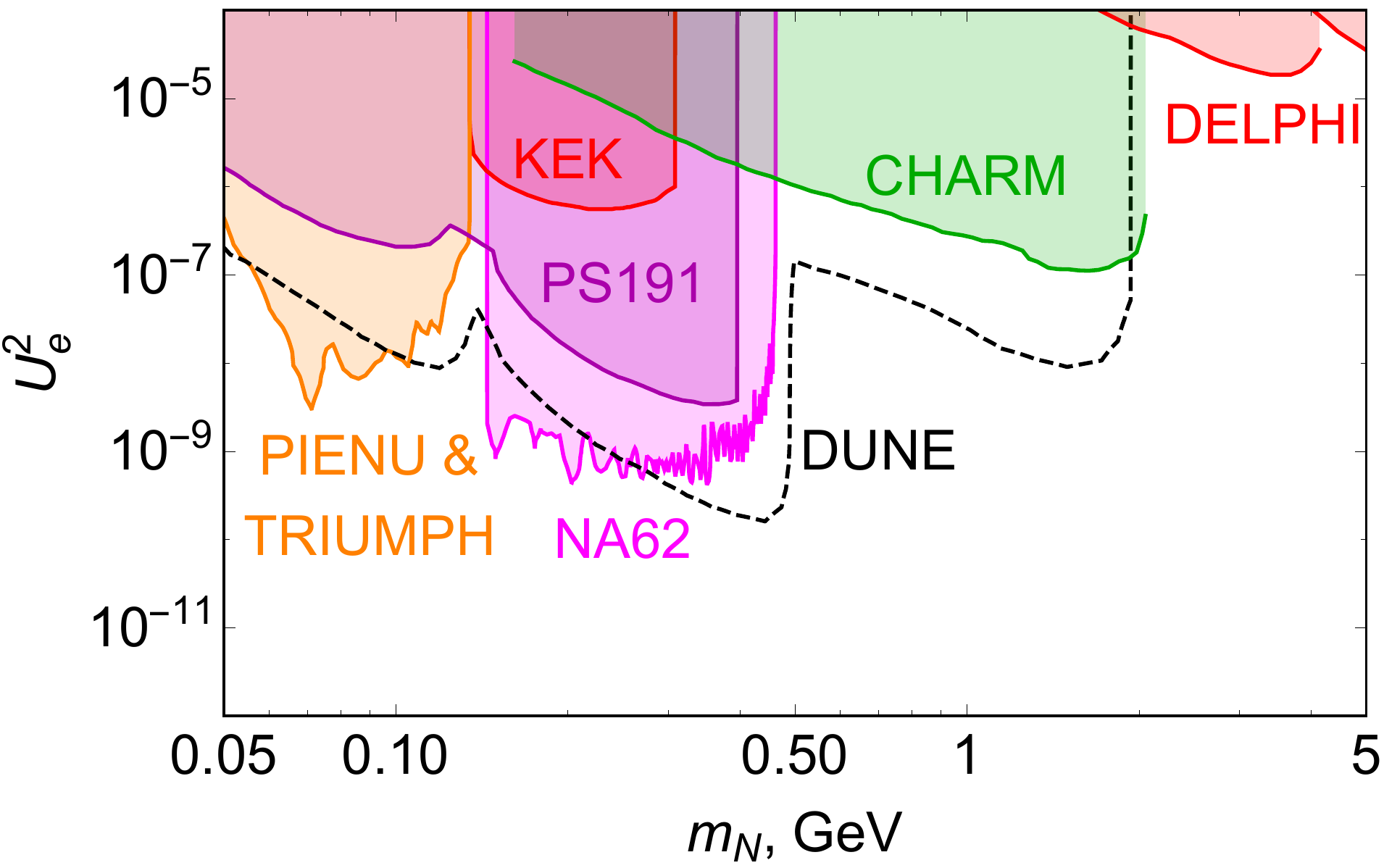"}~\includegraphics[width = 0.47\textwidth]{"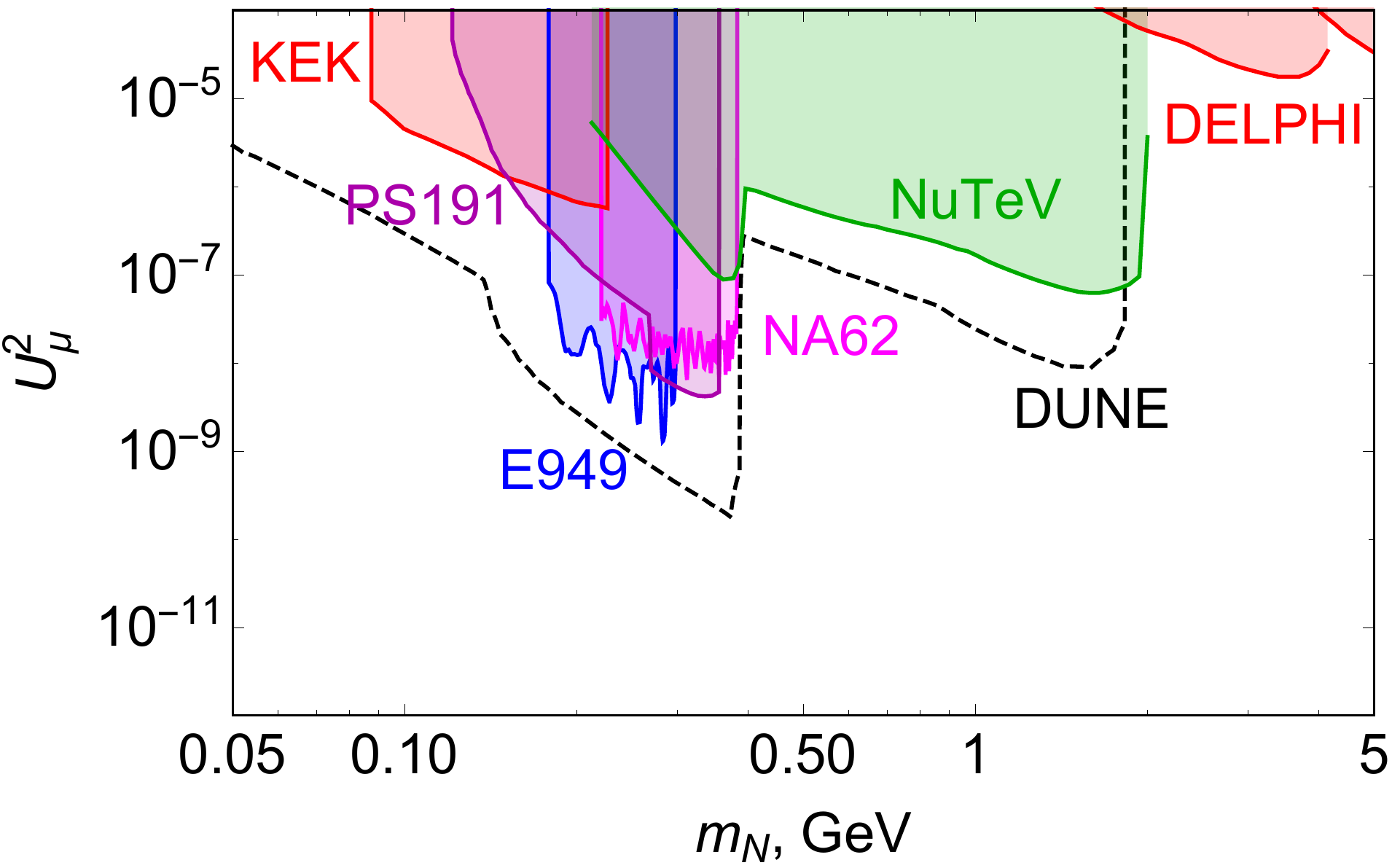"}
	\caption{Accelerator bounds for $U^2_e$ (left panel) and $U^2_\mu$ (right panel) for the HNL mass below $5$~GeV. Also, the expected DUNE sensitivity~\cite{Coloma:2020lgy} is shown (dashed line).}
    \label{fig:UeUmu bounds}
\end{figure}

The second type of experiments is \emph{displaced vertices} search for appearance of SM particles in the decays of long-lived HNLs. This type of experiments can probe combinations $U^2_\alpha U^2_\beta$ because production and decay channels can be governed by different mixing angles. 
The relevant experiments are PS-191~\cite{Bernardi:1985ny, Bernardi:1987ek},\footnote{We note that constraints from PS-191 used here should be considered with caution. Using simple estimates according to~\cite{Bondarenko:2019yob}, we were not able to reproduce the claimed sensitivity of the PS-191 and suspect that the number of kaons in the original analysis was overestimated. However, as constraints from PS-191 do not significantly change the results of our analysis we leave this question for the future analysis.} CHARM~\cite{Bergsma:1985is}, NuTeV~\cite{Vaitaitis:1999wq} as well as  DELPHI~\cite{Abreu:1996pa}. The experimental bounds for pure $U_e^2$ and $U_\mu^2$ mixings are shown in Fig.~\ref{fig:UeUmu bounds}.

\begin{figure}[h!]
	\centering
	\includegraphics[width=0.47\textwidth]{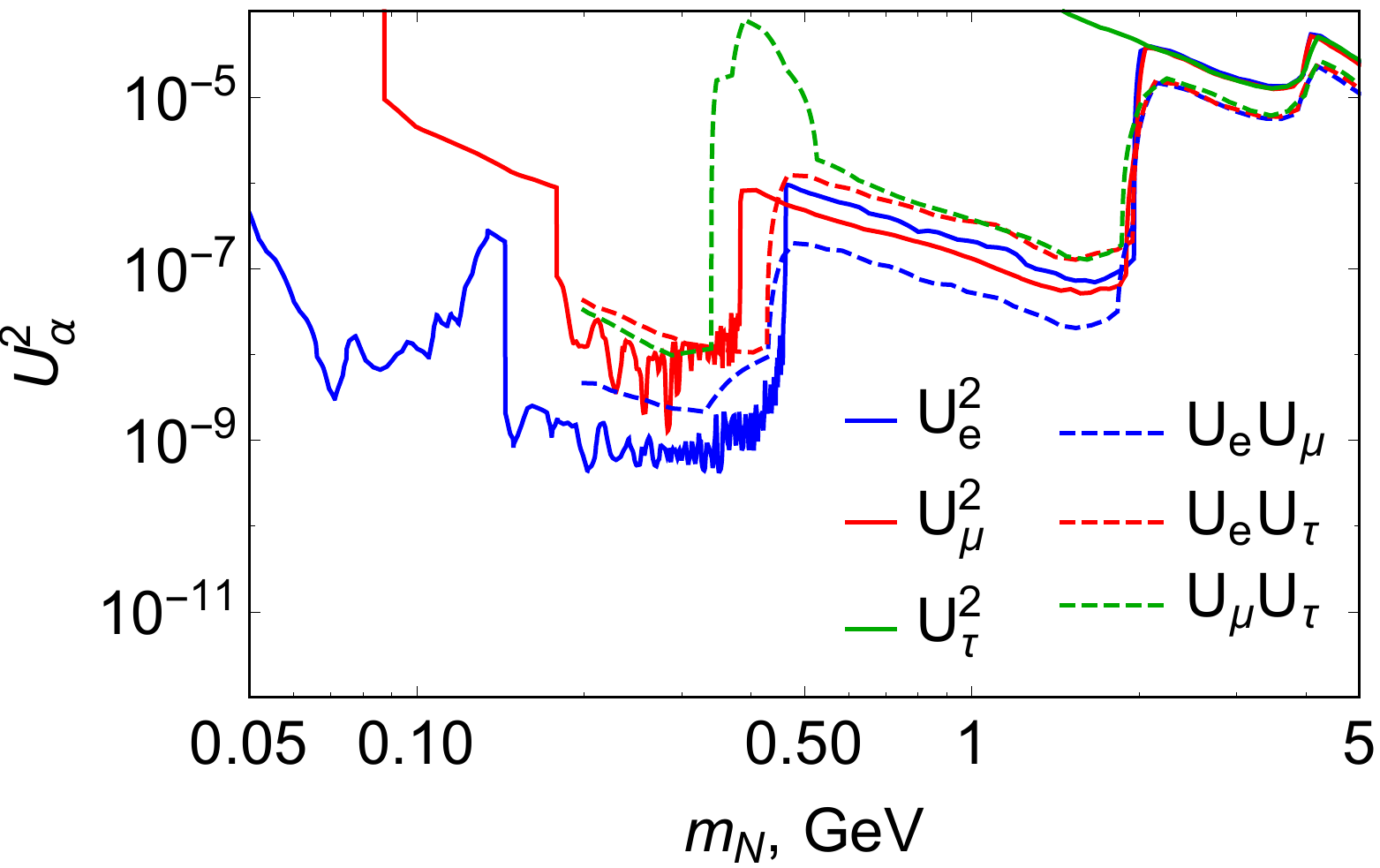}~\includegraphics[width=0.47\textwidth]{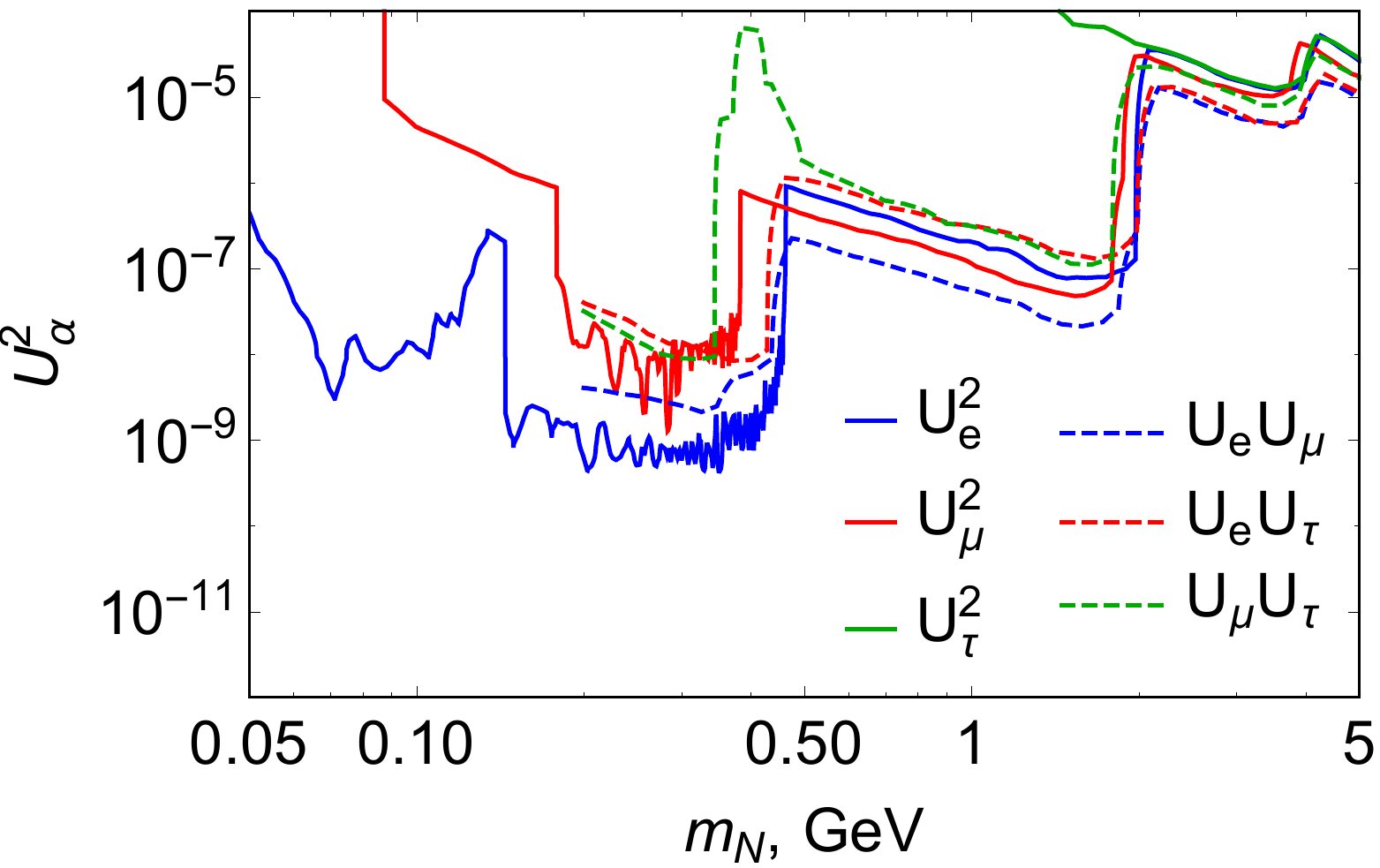}
	\caption{Full set of bounds used in this work for normal hierarchy (left panel) and for inverted hierarchy (right panel). For the $U_\alpha U_{\beta\neq\alpha}$  and $U^2_\tau$ bounds we use only GAMBIT results~\cite{Chrzaszcz:2019inj} starting from $m_N = \unit[0.2]{GeV}$.}
	\label{fig:acc bounds}
\end{figure}

The displaced vertex experiments typically report bounds only on some mixings. 
The reanalysis including bounds was done using GAMBIT in~\cite{Chrzaszcz:2019inj} for the general case of 3 HNLs. We use these results from $m_N>\unit[0.2]{GeV}$ and combine them with results from missing energy experiments. Also, to cover the small window $m_N \approx \unit[0.13-0.14]{GeV}$ in the $U^2_e$ bound we have included explicitly PS-191 results for $U^2_e$ reanalyzed following the prescriptions given in~\cite{Ruchayskiy:2011aa}. The full set of bounds used in this work is shown in Fig.~\ref{fig:acc bounds}.

To combine accelerator limits with other constraints for a given mixing pattern $x_\alpha$ we estimate the actual upper bounds on $U^2$. To find it we need to take into account that $U^2_\tau$ is typically less constrained compared to $U^2_e$, $U^2_\mu$, however large values of $U^2_\tau \gg U^2_{e,\mu}$ (i.e. $x_\tau \approx 1$) are not allowed by neutrino oscillation data. Therefore, for each mixing pattern we compute the maximal mixing angle that does not contradict to any of the $U_\alpha U_\beta$ bounds using:
\begin{equation}
	U^2_{\text{upper}}(x_\alpha) = \min \left(\frac{U^2_{e,\text{acc}}}{x_e} , \frac{(U_{e} U_{\mu})_{\text{acc}}}{\sqrt{x_e x_\mu}}, \frac{U^2_{\mu,\text{acc}}}{x_\mu}, \dots \right)
	\label{eq:acc bounds}
\end{equation}

\section{Constraints from Big Bang Nucleosynthesis}
\label{sec:constraints_from_big_bang_nucleosynthesis}

Accelerator searches provide upper bound on HNL mixing angles~\eqref{eq:acc bounds}. 
On the other hand, a requirement that the presence of HNLs in the primordial plasma would not lead to the over-production of light elements (Deuterium, Helium-4) provides a \textit{lower} bound on the HNL mixing angles.
For HNLs heavier than $\pi^\pm$-mesons the  strongest BBN bound of the HNL lifetime is due to $n\leftrightarrow p$ meson driven conversion~\cite{Boyarsky:2020dzc}. Pions and kaons produced in HNL decays at the time when free neutrons are present in the plasma modify the resulting freeze-out ratio of neutron to proton abundances, leading to a larger values of $\isotope[4]{He}$ abundance as compared to the Standard Model BBN. If meson production is kinematically allowed, the following constraint can be derived~\cite{Boyarsky:2020dzc}:
\begin{equation}
	\label{eq:BBN meson}
	\tau_N \lesssim \frac{\unit[0.023]{s}
	}{1 + 0.07 \ln\left( \dfrac{P_{\text{conv}}}{0.1} \dfrac{\text{Br}_{N\rightarrow h}}{0.1} \dfrac{Y_N \zeta}{10^{-3}}\right)}\,,
\end{equation}
where $P_\text{conv}$ is the probability for meson to interact before decaying, $\text{Br}_{N\rightarrow h}$ is the branching fraction of semileptonic HNL decays producing a given meson $h$, $Y_N$ is the initial HNL abundance, and $\zeta\equiv (a_{\text{SM}}/a_{\text{SM+HNLs}})^{3}<1$ is the dilution factor. In the combination $P_{\text{conv}} \text{Br}_{N\rightarrow h}$ a summation over meson species is assumed. Note the logarithmic dependence on these parameters, since for $\tau_N\ll \unit[0.1]{s}$ HNLs and consequently mesons have exponentially small abundances at the time of interest.

Implementation of different mixing patterns changes the value of $\text{Br}_{N\rightarrow h}$ only, since $Y_N \zeta$ depends on processes at high temperature, where all lepton species are in equilibrium, and $P_{\text{conv}}$ is solely related to mesons.  The value of $Y_N \zeta$ varies in $10^{-3} - 10^{-2}$, therefore we use the conservative lower bound $Y_N \zeta = 10^{-3}$~\cite{Boyarsky:2020dzc}. In terms of $(U^2_e, U^2_\mu, U^2_\tau) = U^2 (x_e, x_\mu, x_\tau)$, the branching ratio can be parametrized in the following way:
\begin{equation}
	\text{Br}_{N\rightarrow h} = \sum_{X \in \text{states with }h} n_h(X) \frac{x_e \Gamma(N_e \rightarrow X) + x_\mu \Gamma(N_\mu \rightarrow X) + x_\tau \Gamma(N_\tau \rightarrow X)}{x_e \Gamma(N_e) + x_\mu \Gamma(N_\mu ) + x_\tau \Gamma(N_\tau)}
\end{equation}
where the notation $N_\alpha$ corresponds to an HNL with the mixing angles $U^2_\alpha = 1$ and $U^2_{\beta\neq \alpha} = 0$, $\Gamma(N_\alpha)$ is the total decay width, $\Gamma(N_\alpha \rightarrow X)$ is the HNL decay width into state $X$, and $n_h(X)$ is the meson $h$ multiplicity for the final state $X$. For a given mixing pattern it is straightforward to compute the corresponding $P_{\text{conv}} \text{Br}_{N\rightarrow h}$ and substitute in \eqref{eq:BBN meson}.

\begin{figure}[t]
	\centering
	\includegraphics[width = 0.5\textwidth]{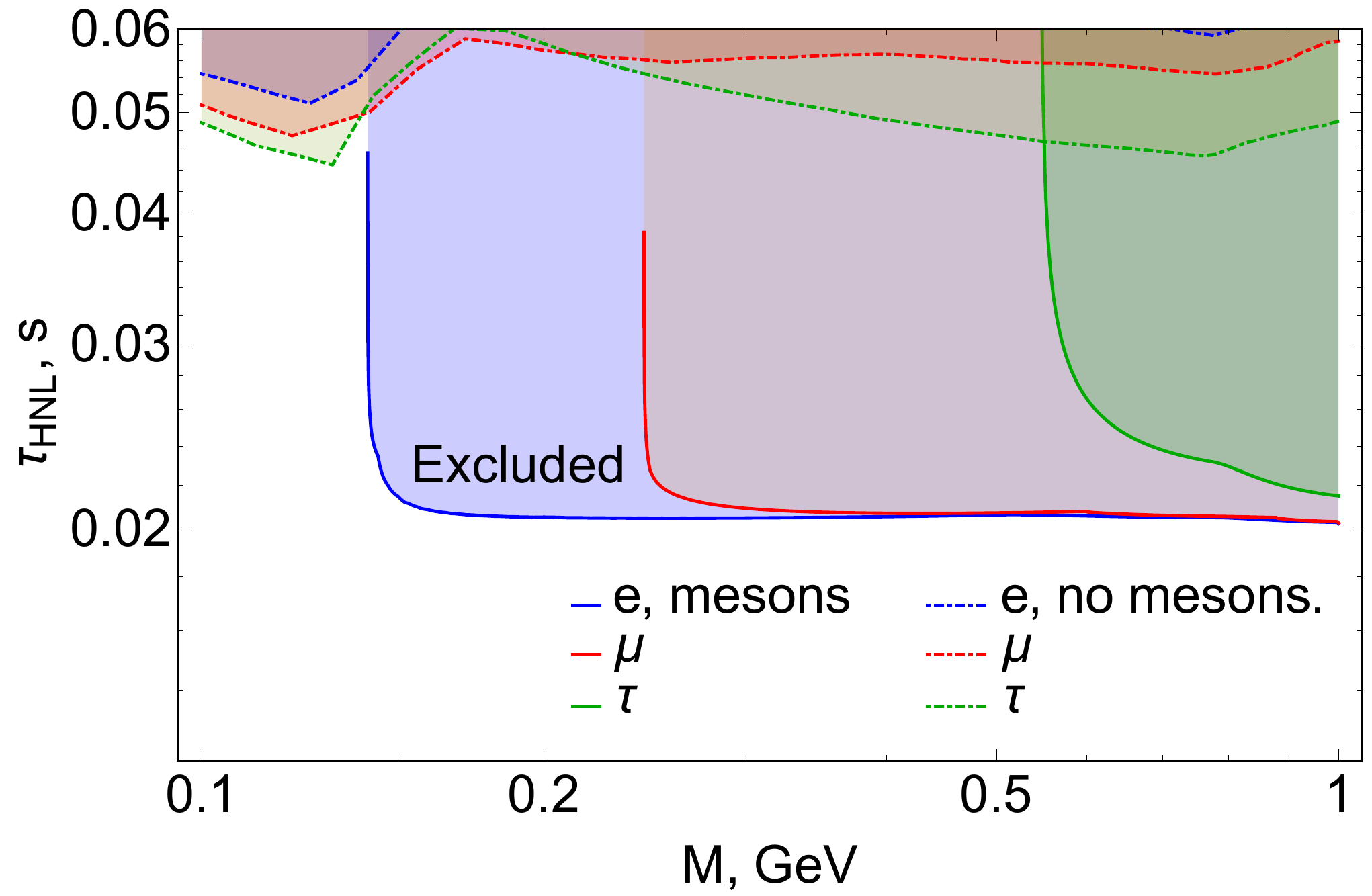}
	\includegraphics[width = 0.47\textwidth]{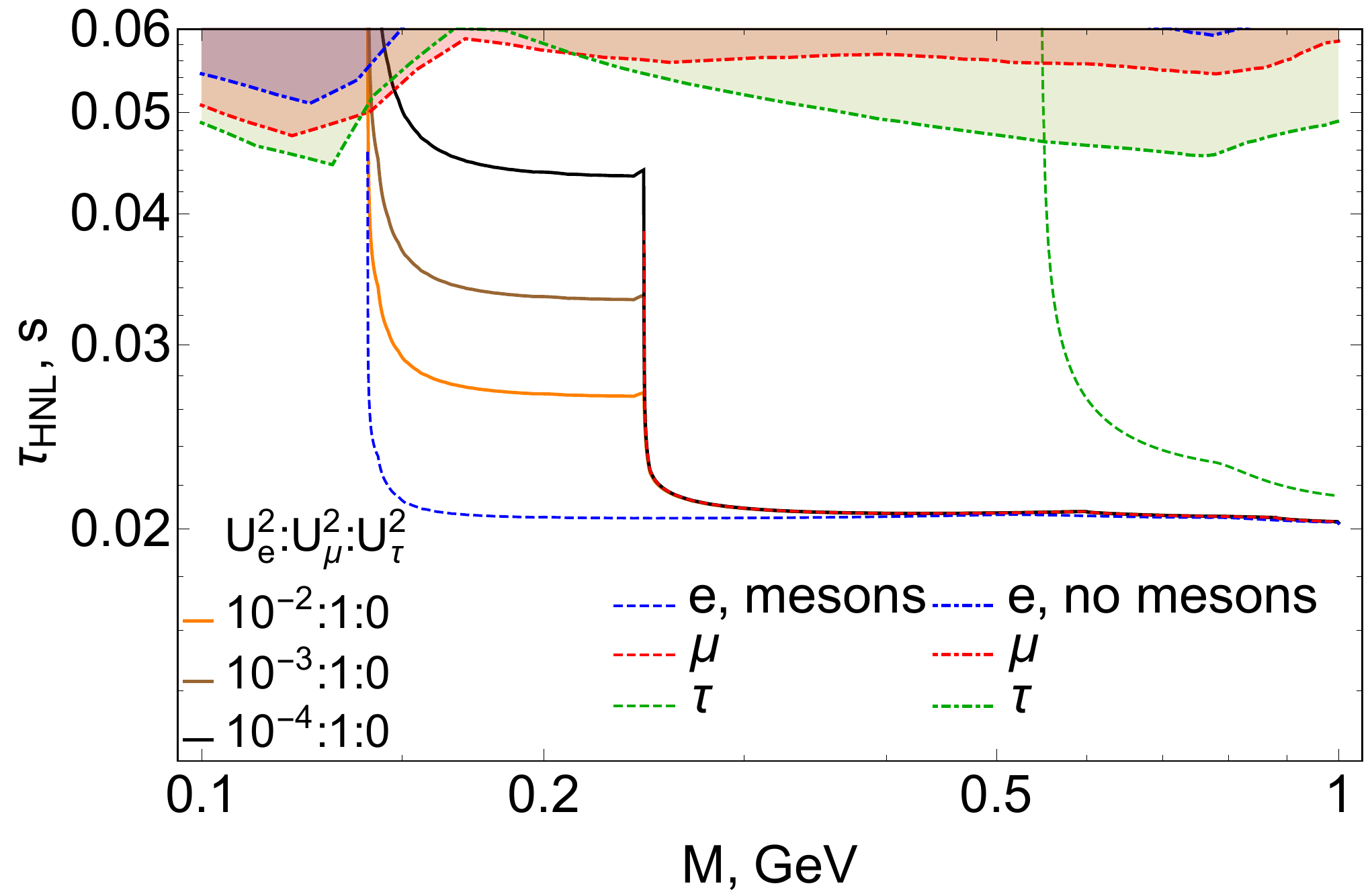}~\includegraphics[width = 0.47\textwidth]{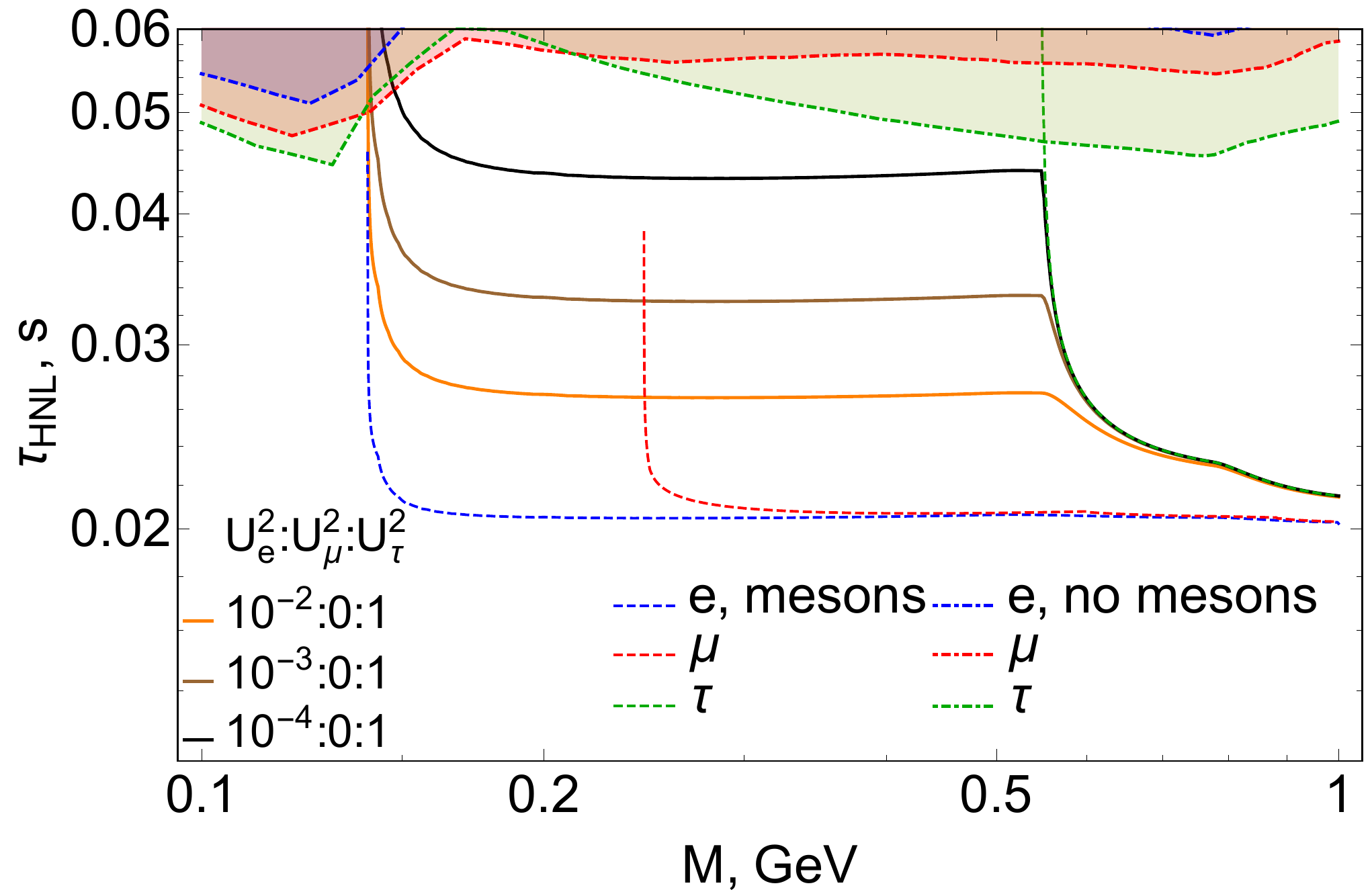}
	\caption{\textit{Top panel}: the lifetime bounds for the pure mixing cases. \textit{Left panel}: muon mixing with a small contribution of $U^2_e$. \textit{Right panel}: tau mixing with a small contribution of $U^2_e$.}
	\label{fig:BBNlimits}
\end{figure}

For pure mixing cases the bound is applicable only for HNL masses exceeding meson production threshold:
\begin{align*}
	&m_N > m_\pi + m_e \approx \unit[130]{MeV}& &\text{for electron mixing,}&\\
	&m_N > m_\pi + m_\mu \approx \unit[240]{MeV}& &\text{for muon mixing,}&\\
	&m_N > m_\eta \approx \unit[550]{MeV} & &\text{for tau mixing.}&
\end{align*}
However, even small fraction of $U^2_e$ can relax this restriction to $m_N > m_\pi+m_e$ due to logarithmic dependence on the total branching ratio, see examples in Fig.~\ref{fig:BBNlimits}.

For the parameter region where the meson constraint does not work we use a conservative estimate $\tau_N < \unit[0.1]{s}$ from~\cite{Sabti:2020yrt}.\footnote{The actual estimate on the HNL lifetime depends on the maximally admissible value of 
$\Delta Y_{\text{He}}/Y_{\text{He}}$. Here we use $\Delta Y_{\text{He}}/Y_{\text{He}}\leq 4.35\%$  used in~\cite{Boyarsky:2020dzc} and adopted in this work. Ref.~\cite{Sabti:2020yrt} refers a twice stronger bound, because it adopts tighter margin for $Y_{\text{He}}$.} 
Taking this into account the resulting  expression for the lower bound for $U^2$ is
\begin{equation}
	U^2_{\text{lower}}(x_\alpha) = \frac{1}{\tau^{\text{BBN}}_{N} (x_\alpha) \cdot \sum_\alpha x_{\alpha} \Gamma(N_\alpha)}
\end{equation}
where $\tau^{\text{BBN}}_N$ is given by the minimal value between the r.h.s. of \eqref{eq:BBN meson} and $\unit[0.1]{s}$.

\section{Constraints from Leptogenesis}
\label{sec:constraints_from_leptogenesis}
The smallness of the light neutrino masses is not the only problem HNLs can solve, they can also explain the observed BAU through leptogenesis~\cite{Fukugita:1986hr}.
The condition of reproducing the observed BAU~\cite{Aghanim:2018eyx,Zyla:2020zbs},
\begin{align}
    \frac{n_B}{n_\gamma}= (5.8 - 6.5) \times 10^{-10}\,,
\end{align}
imposes further constraints on the properties of the HNLs.
When combined with the bounds from the seesaw mechanism, leptogenesis imposes a strong constraint on mass spectrum of the HNLs, namely it forbids hierarchical HNL masses if the lightest mass is below $10^9$ GeV~\cite{Davidson:2002qv}.
This implies that (in the minimal model) any HNLs we can observe in the near future are degenerate in mass,
and that leptogenesis is realized either via a resonant 
enhancement in HNL decays~\cite{Liu:1993tg,Flanz:1994yx,Flanz:1996fb,Covi:1996wh,Covi:1996fm,Pilaftsis:1997jf,Pilaftsis:1997dr,Pilaftsis:1998pd,Buchmuller:1997yu,Pilaftsis:2003gt},
or via HNL oscillations~\cite{Akhmedov:1998qx,Asaka:2005pn}.
If we combine these two mechanisms, leptogenesis is possible for all HNL masses larger than $\sim 100$ MeV~\cite{Klaric:2020lov}.

\begin{figure}[h!]
	\centering
	\includegraphics[width = 0.49\textwidth]{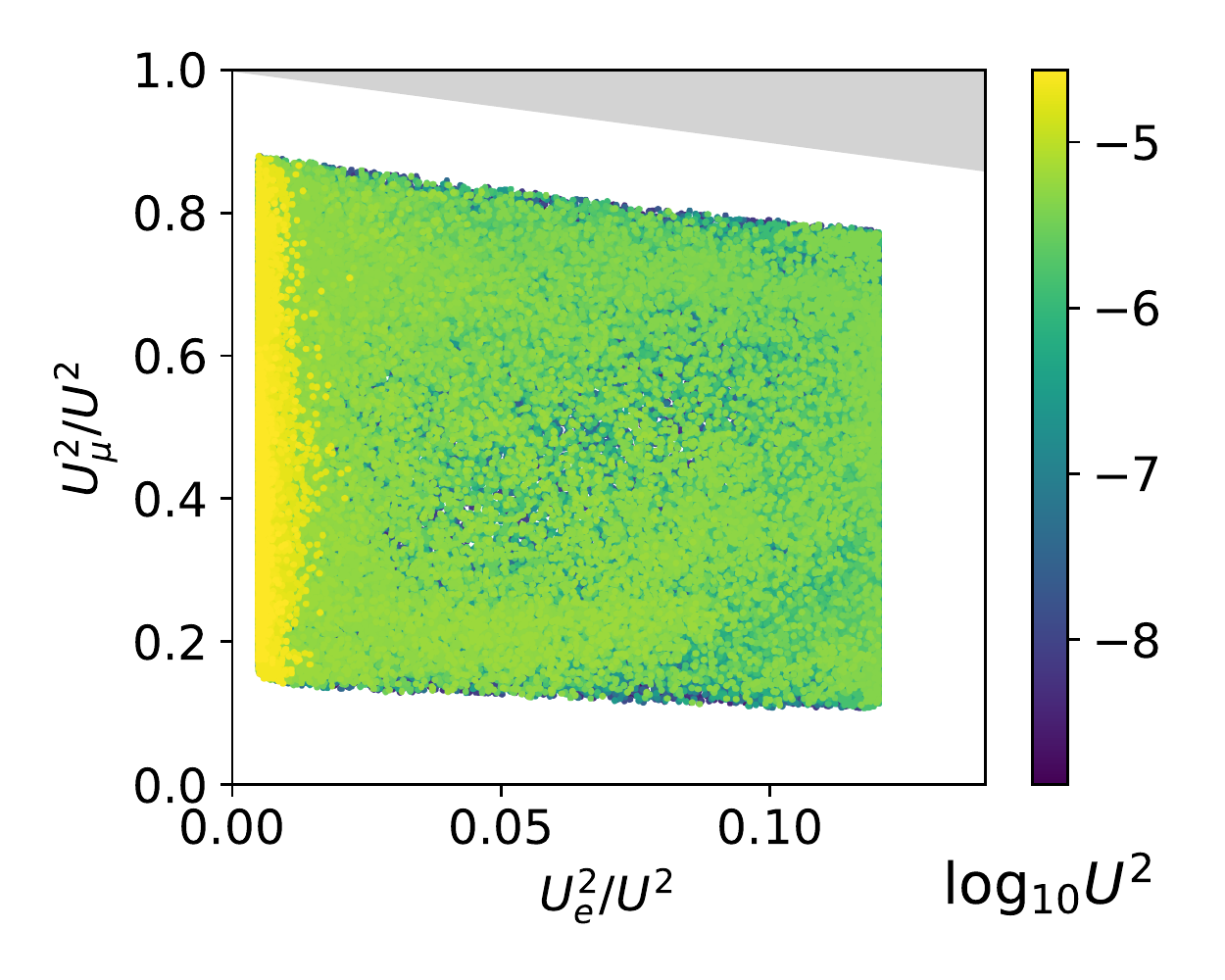}
	\includegraphics[width = 0.49\textwidth]{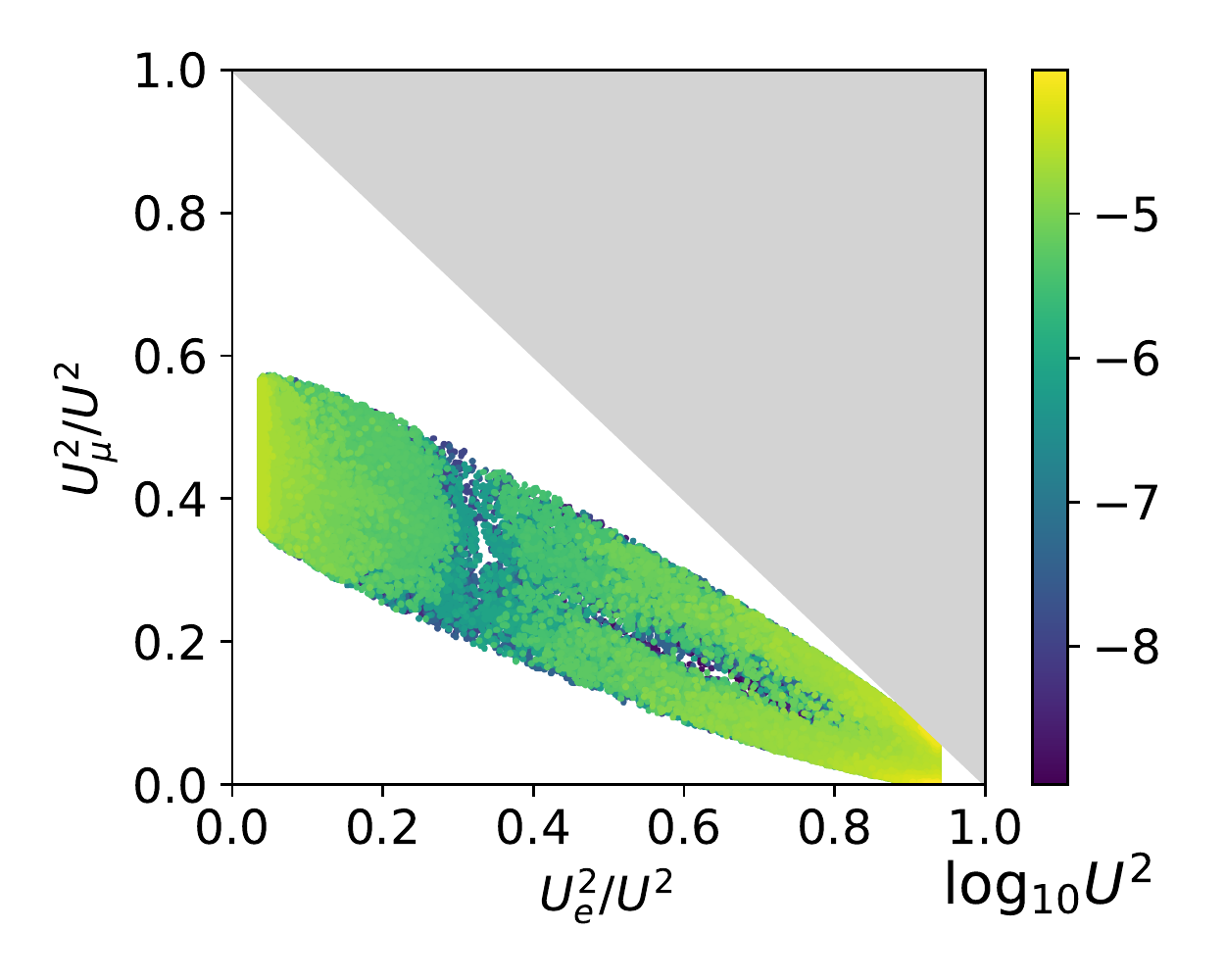}
	\caption{
	Flavor patterns consistent with both the neutrino oscillation data (as in Fig.~\ref{fig:oscillation-bounds}), and leptogenesis for $M_N \sim \unit[140]{MeV}$. The upper bound on $U^2$ depends on the ratios $U_\alpha^2/U^2$, as this can prevent a large washout of the lepton asymmetries. The color coding indicates the maximal $U^2$ for which baryogenesis via leptogenesis remains possible.
	We note here that the experimental and BBN bounds on the mixing angles are not included in these figures, as in this range of HNL masses they completely dominate over the constraints from leptogenesis, as shown in Fig.~\ref{fig:140mev}.
	}
	\label{fig:lg_flv}
\end{figure}

Nonetheless, leptogenesis can also provide other interesting constraints on the HNL properties.
Phenomenologically, the most important constraint is the limit on the maximal size of the HNL mixing angles $U^2$~\cite{Shaposhnikov:2008pf,Canetti:2012kh,Drewes:2016gmt,Hernandez:2016kel,Drewes:2016jae,Antusch:2017pkq,Eijima:2018qke,Klaric:2020lov}.
This limit arises from the fact that for large mixing angles the HNL interactions become too fast, and the lepton number reaches thermal equilibrium before the sphalerons freeze-out at $T\sim130$ GeV.
\paragraph{Allowed flavor mixing patterns.}
The upper bounds on $U^2$ can have a strong dependence on the choice of flavor mixing pattern~\cite{Drewes:2016jae,Antusch:2017pkq,Eijima:2018qke}, as shown in Fig.\ \ref{fig:lg_flv}.
A tiny mixing with a particular lepton flavor means that lepton flavor will equilibrate more slowly in the early Universe, and can thus prevent complete equilibration of lepton number.
The allowed mixing patterns are almost completely determined by the low-energy phases as shown in Fig~\ref{fig:oscillation-bounds}.
This means that the leptogenesis bounds can also shift as the neutrino oscillation data is updated.
For example, in the case of inverted hierarchy, the choice of optimal phases corresponded to $\delta_{\text{CP}} = 0$~\cite{Drewes:2016jae,Eijima:2018qke}, which is disfavored by the latest fits of the light neutrino parameters~\cite{Esteban:2020cvm}.

\begin{figure}[t!]
    \centering
    \includegraphics[width = 0.6\textwidth]{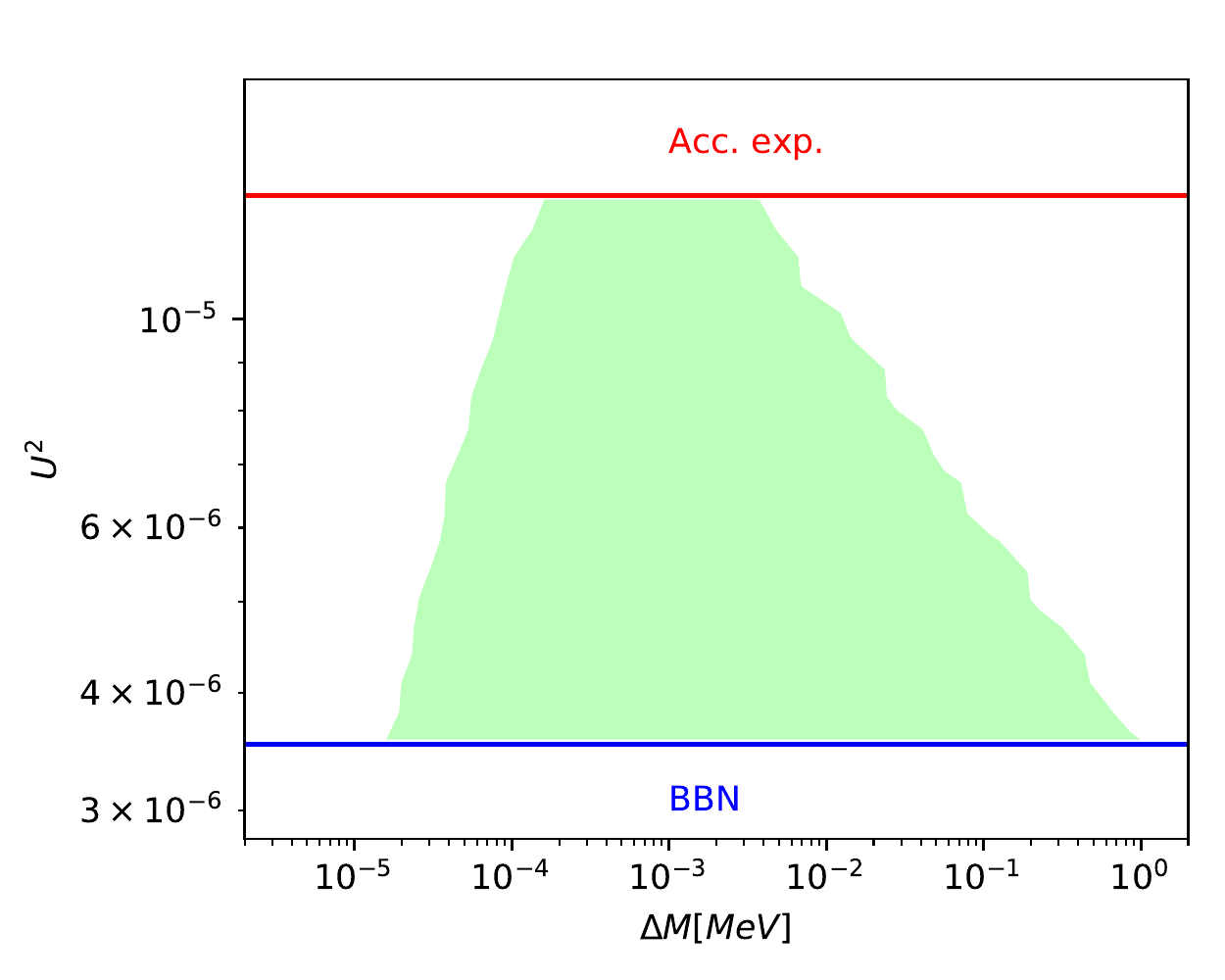}
    \caption{The allowed range of HNL mass splittings consistent with leptogenesis for a benchmark mass $M_N=\unit[140]{MeV}$.
    All points are consistent with the experimental constraints. Interestingly, relatively large $(\unit[1]{MeV})$ mass splittings are allowed, which could potentially be resolved at experiments. It is interesting that all mass splittings are large enough that the rates of lepton number violating and conserving decays are approximately equal.
    }
    \label{fig:dMU2tau}
\end{figure}

\paragraph{The HNL Mass splitting.}
The mass splitting between the HNLs is one of the key parameters determining the size of the BAU.
The condition of successful leptogenesis constrains the maximal size of the mass splitting (see, e.g.~\cite{Canetti:2012kh,Drewes:2016gmt,Hernandez:2016kel,Drewes:2016jae,Antusch:2017pkq,Eijima:2018qke,Klaric:2020lov}),
which can have direct consequences for the various lepton number violating signatures at direct search experiments~\cite{Cvetic:2015ura,Anamiati:2016uxp,Gluza:2015goa,Dev:2015pga,Antusch:2017ebe,Drewes:2019byd,Tastet:2019nqj,Antusch:2020pnn}, or for the indirect signatures such as neutrinoless double beta decay~\cite{Drewes:2016lqo,Hernandez:2016kel}.
As an example, for $M_N \approx 140$ MeV,
we show how leptogenesis constrains the remaining parameters after we apply all the other cuts.
The allowed range of mass splittings $\Delta M_N = |M_2-M_1|/2$, consistent with leptogenesis depends on $U^2$ and is shown in Fig.~\ref{fig:dMU2tau}.

\section{Results}
\label{sec:combined-limits}
\subsection{Numerical procedure}
\label{sec:procedure}
Our procedure of finding viable HNLs models (green points) is as follows.
We consider two HNLs  degenerate in mass that pass \emph{all of the following constraints}:
\begin{enumerate}[1.]
    \item \textbf{The mixing angles $U_\alpha^2(x)$ are chosen such that neutrino oscillation data is satisfied.} This is ensured by the Casas-Ibarra parametrization~\eqref{eq:mixing pattern casasibarra}. By varying the CP phase $\delta_{\text{CP}}$ and $\theta_{23}$ within their $95\%$ confidence region ($\Delta \chi^2 <6.0$) and by changing the unconstrained Majorana phase $\eta \in [0, 2\pi)$ we determine the region of parameters $(x_e, x_\mu)$ admissible by the neutrino oscillation data.
    \item \textbf{All $U_\alpha^2(x)$ must be \emph{smaller} than the corresponding accelerator limits for the flavor~$\alpha$.} To ensure this we scan over the points in the $(x_e, x_\mu)$ plane consistent with neutrino oscillation data and for each mass $M_N$ compute the upper bound $U^2(x)$ from the accelerator experiments
		\begin{equation}
		    \label{eq:U2upper}
			U^2_{\text{upper}}(x_\alpha) = \min \left(\frac{U^2_{e,\text{max}}}{x_e} , \frac{(U_{e} U_{\mu})_{\text{max}}}{\sqrt{x_e x_\mu}}, \frac{U^2_{\mu,\text{max}}}{x_\mu}, \dots \right)
			\end{equation}
			The admissible mixing angles $U_\alpha^2$ should be below $x_\alpha  U^2_{\text{upper}}(x_\alpha)$.
    \item \textbf{All $U_\alpha^2(x)$ must be \emph{larger} than the corresponding BBN bounds for the given flavor.}
    To this end we find
    \begin{equation}
    \label{eq:U2lower}
			U^2_{\text{lower}}(x_\alpha) = \frac{1}{\tau^{\text{BBN}}_{N} (x_\alpha) \cdot \sum_\alpha x_{\alpha} \Gamma(N_\alpha)}
		\end{equation}
		(where the quantities in Eq.~\eqref{eq:U2lower} are defined in Section~\ref{sec:constraints_from_big_bang_nucleosynthesis}) and compute admissible $U_\alpha^2(x) = x_\alpha U^2_{\text{lower}}(x)$ 
	\item All \textbf{$U^2_\alpha$ are minimized/maximized independently} with respect to $x_\alpha$.
		
    \item When we start to approach the seesaw line ($M U_\alpha^2 \approx \sum_i m_i$) two HNLs may in principle have different mixing angles, i.e.\ $U_{\alpha 1} \neq U_{\alpha 2}$ and, correspondingly, different lifetimes.  
    To probe the region near the seesaw bound, i.e.\ when $U^2 M \sim (m_1 + m_2)$ or, equivalently, when $2\operatorname{Im}\omega \to 0$ we used the general expression \eqref{CI:Yukawa} to generate a large sample of points in the range $2\operatorname{Im}\omega \in [0,\ln 100], \operatorname{Re}\omega \in [0,2\pi)$ to
     ensure that the above conditions are satisfied by each of the HNLs.
     
\item Finally \textbf{we ensure that the observed value of BAU can be reproduced}. To this end we  numerically solve the quantum kinetic equations of ref.~\cite{Klaric:2020lov}. 
\end{enumerate}

\subsection{The space of viable 2 HNL models}
\label{sec:viable_models}

\begin{figure}[h!]
	\centering
	\includegraphics[width = 0.49\textwidth]{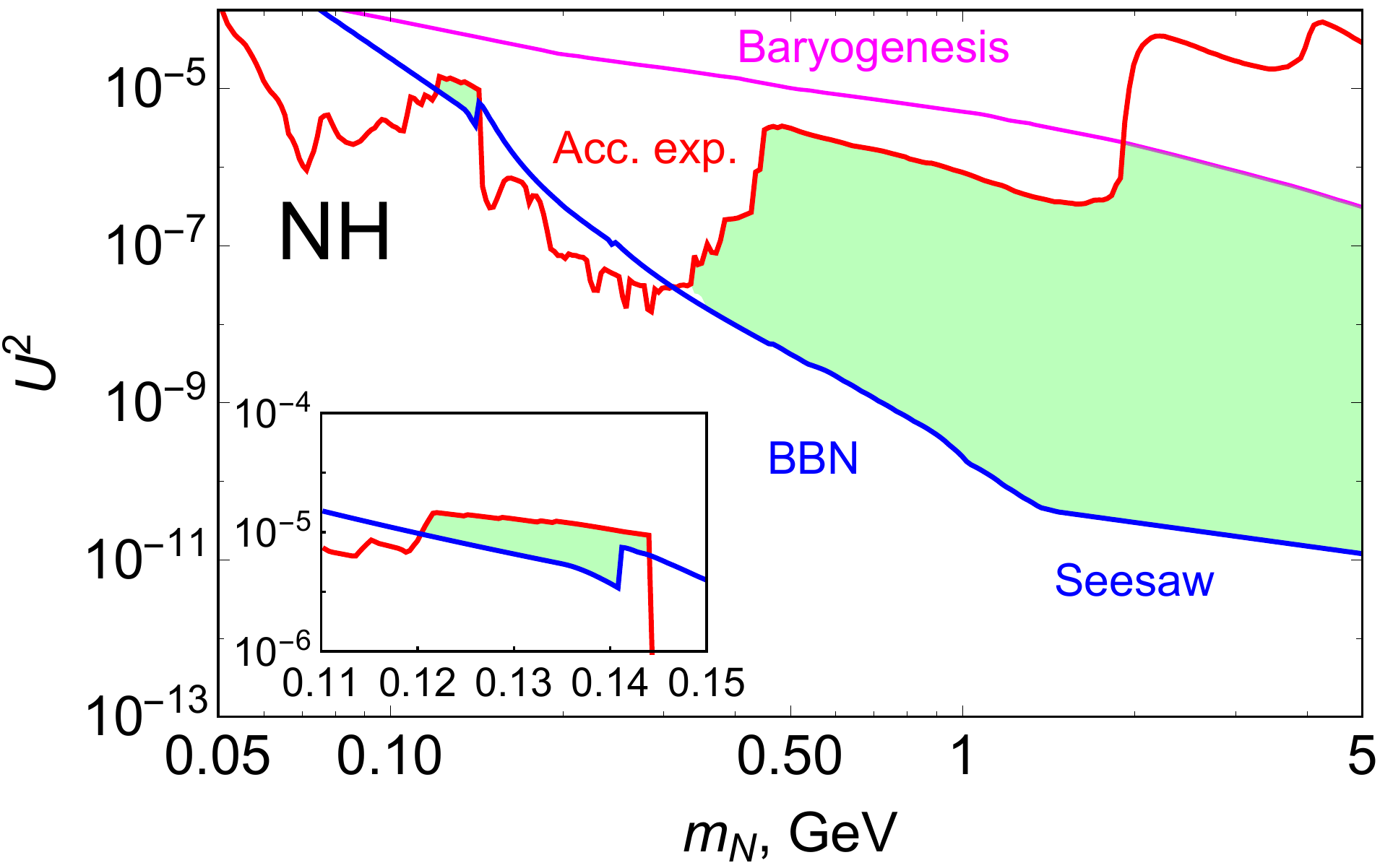}~\includegraphics[width = 0.49\textwidth]{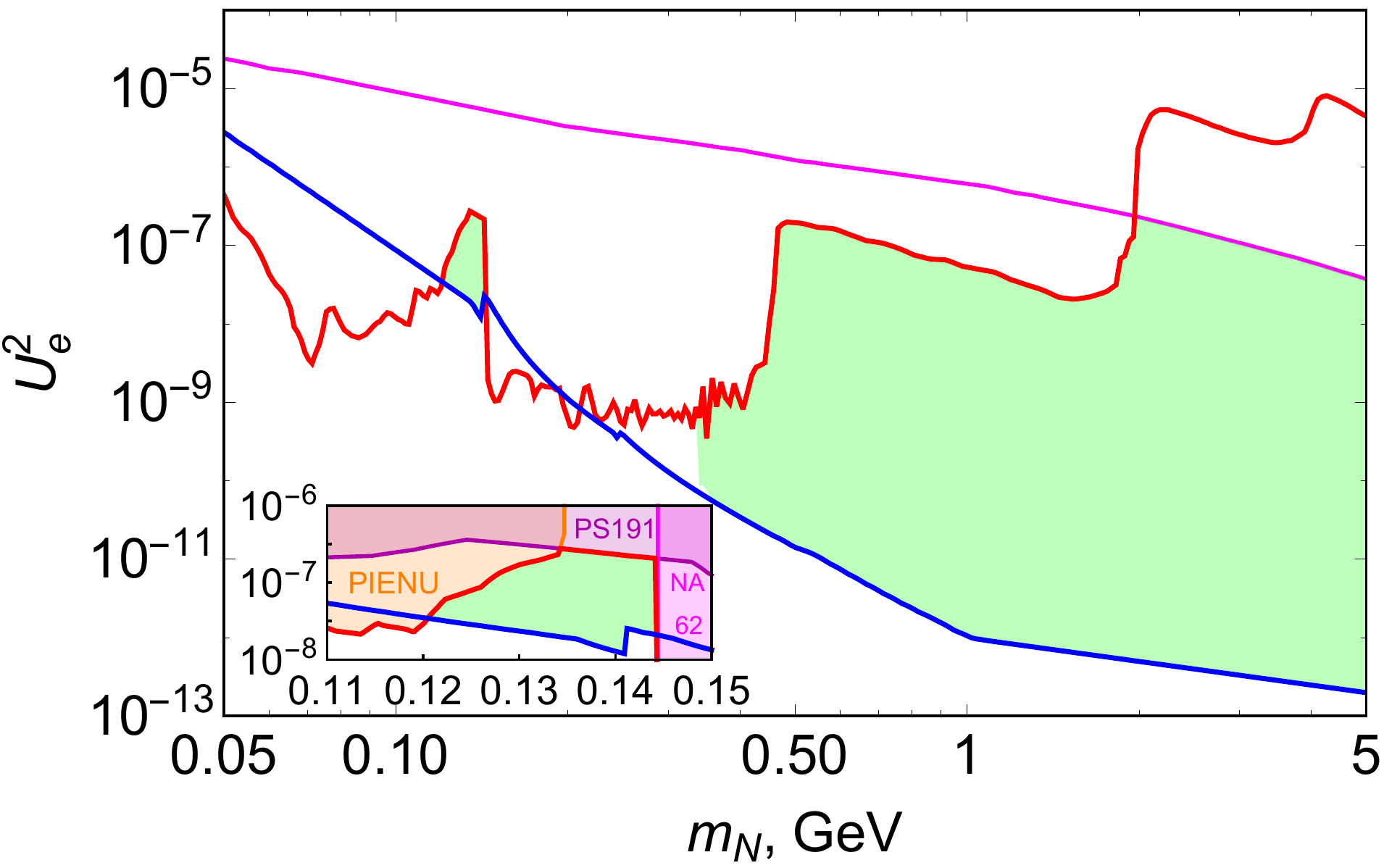}

	\includegraphics[width = 0.49\textwidth]{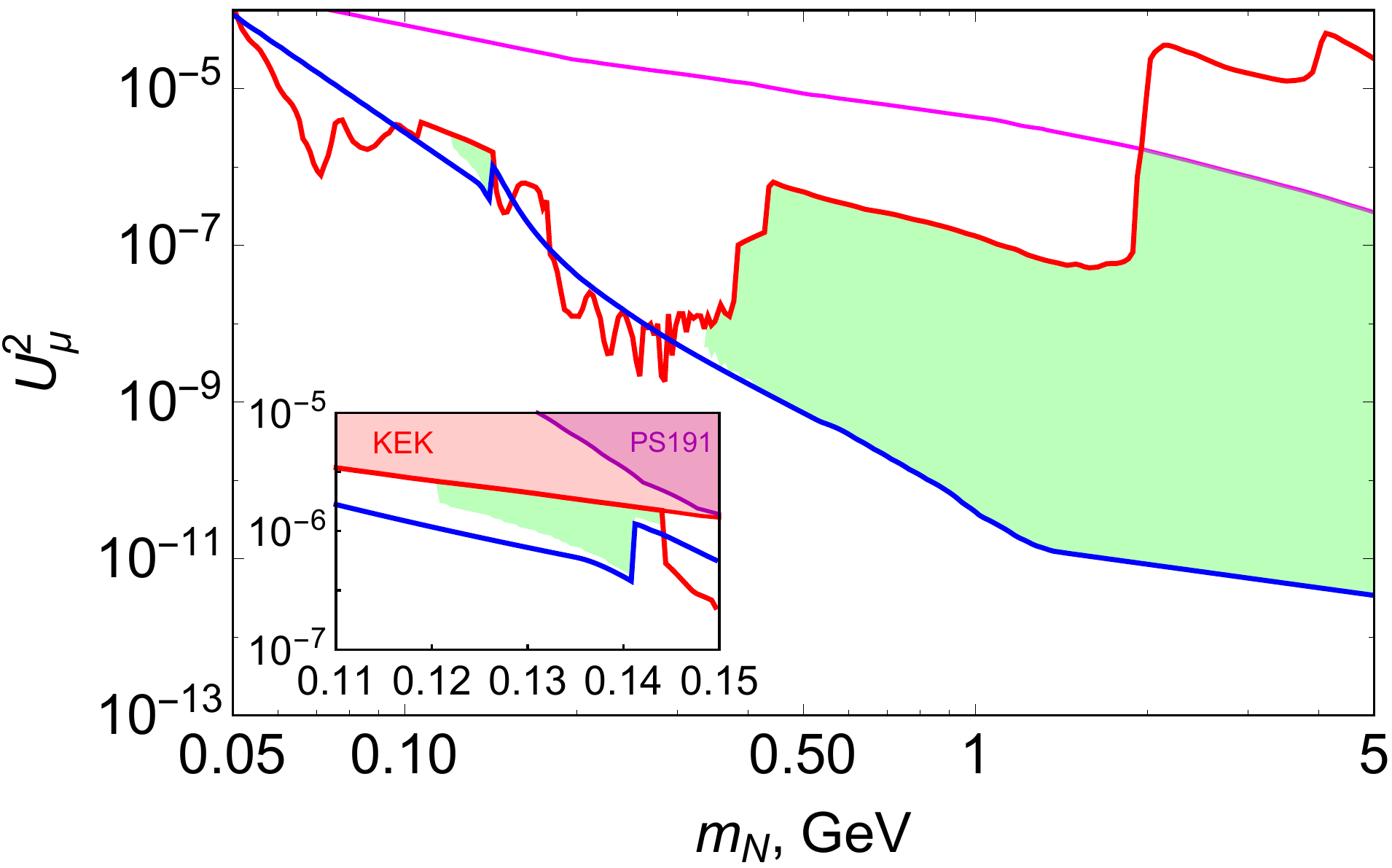}~\includegraphics[width = 0.49\textwidth]{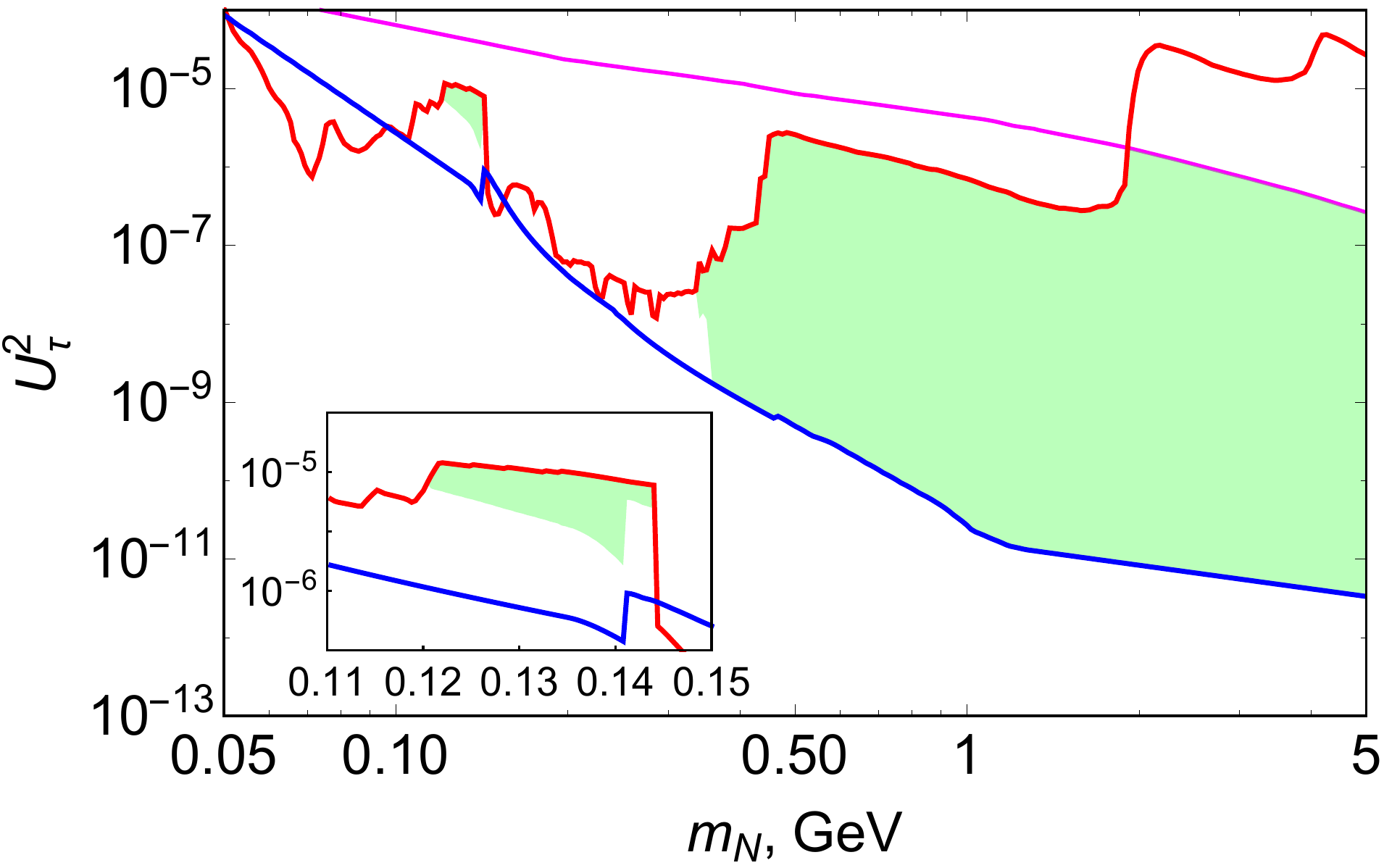}
	\caption{The parameter space of the model with two HNLs. Green points are consistent with all experimental bounds, explain neutrino data for the normal neutrino mass hierarchy (NH) and generate the correct BAU.
	Independent bounds for each flavor from the accelerator experiments (red) and BBN (blue) are also shown.}
	\label{fig:NH result}
\end{figure}

\begin{figure}[h!]
	\centering
	\includegraphics[width = 0.47\textwidth]{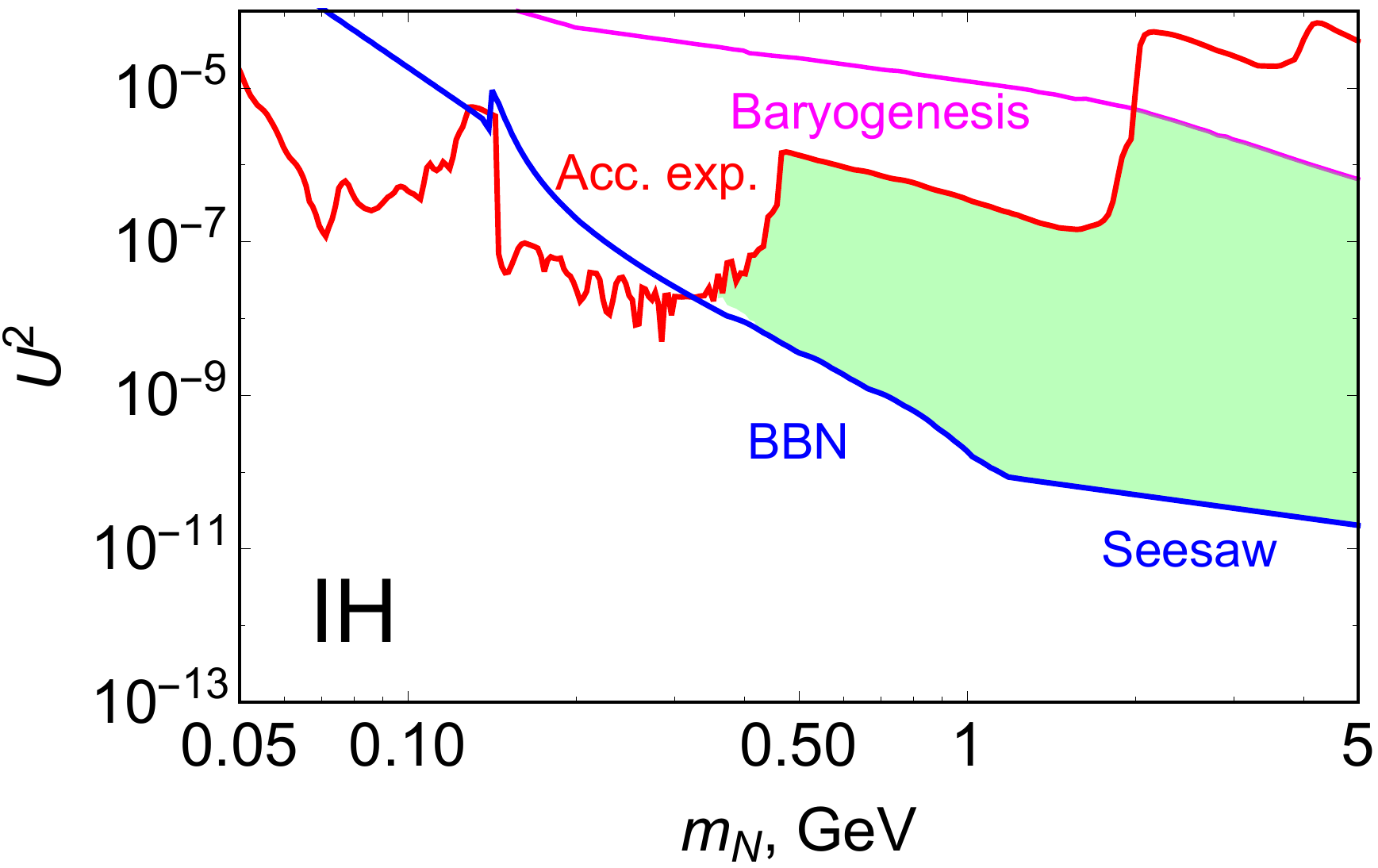}~\includegraphics[width = 0.47\textwidth]{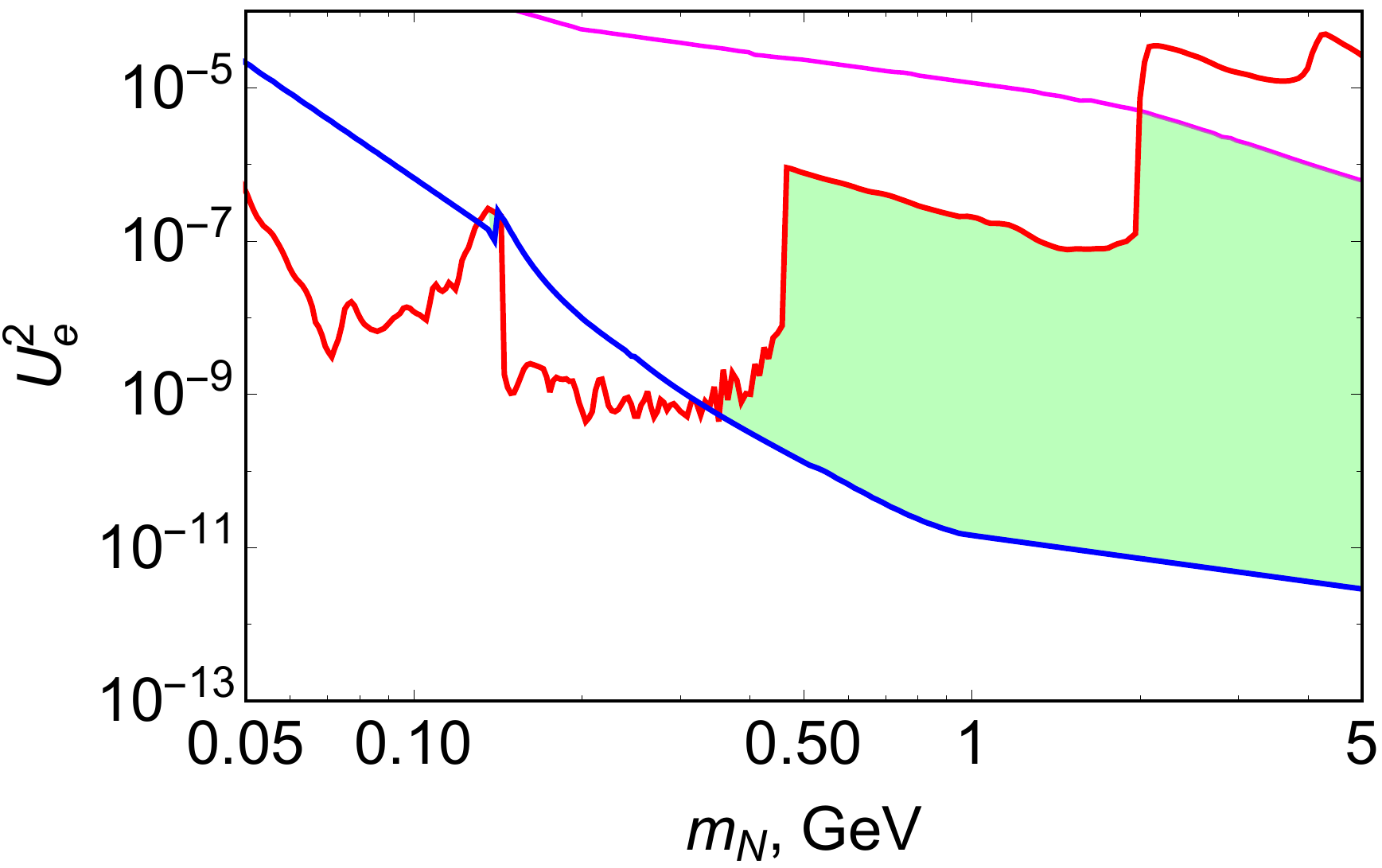}

	\includegraphics[width = 0.47\textwidth]{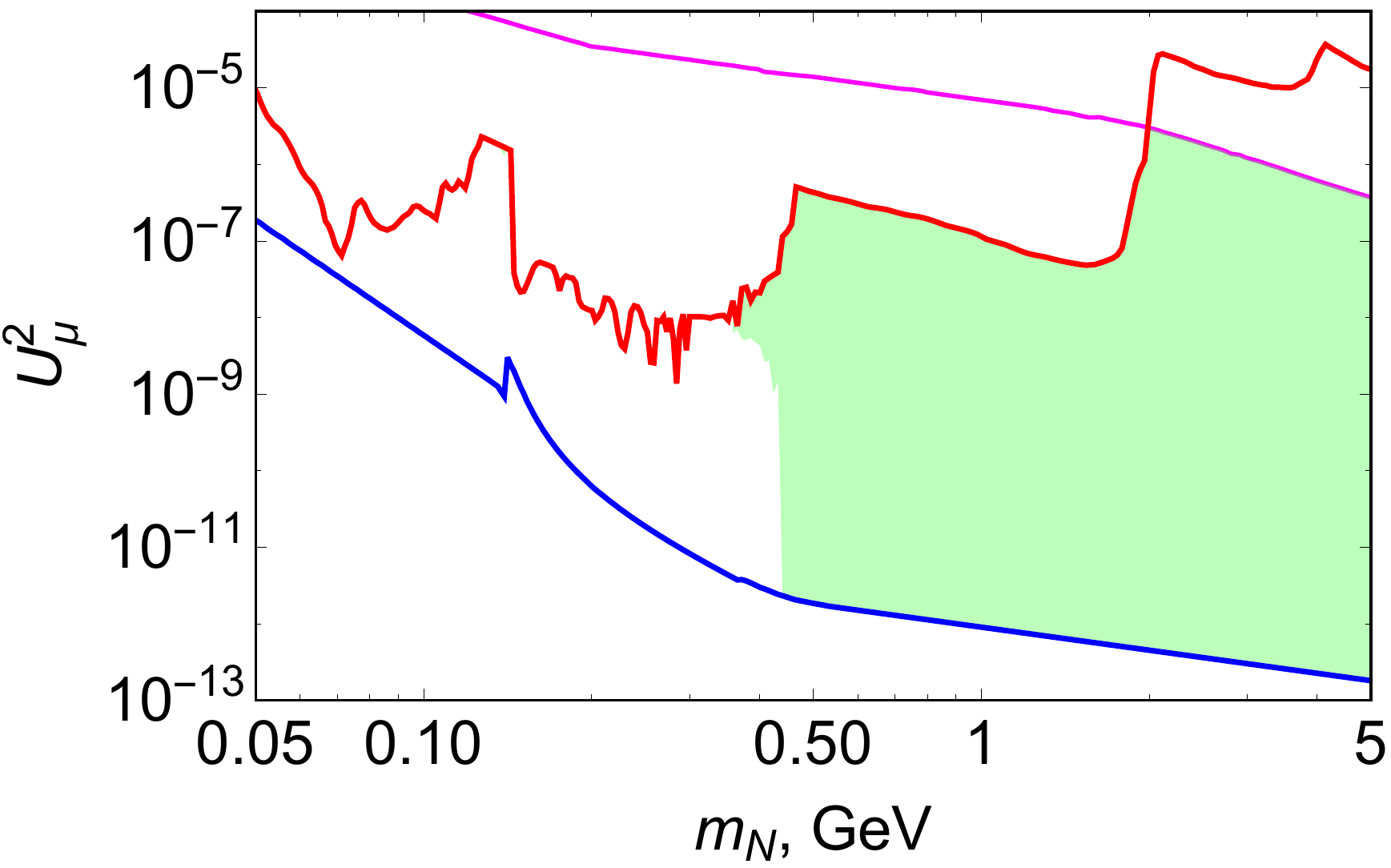}~\includegraphics[width = 0.47\textwidth]{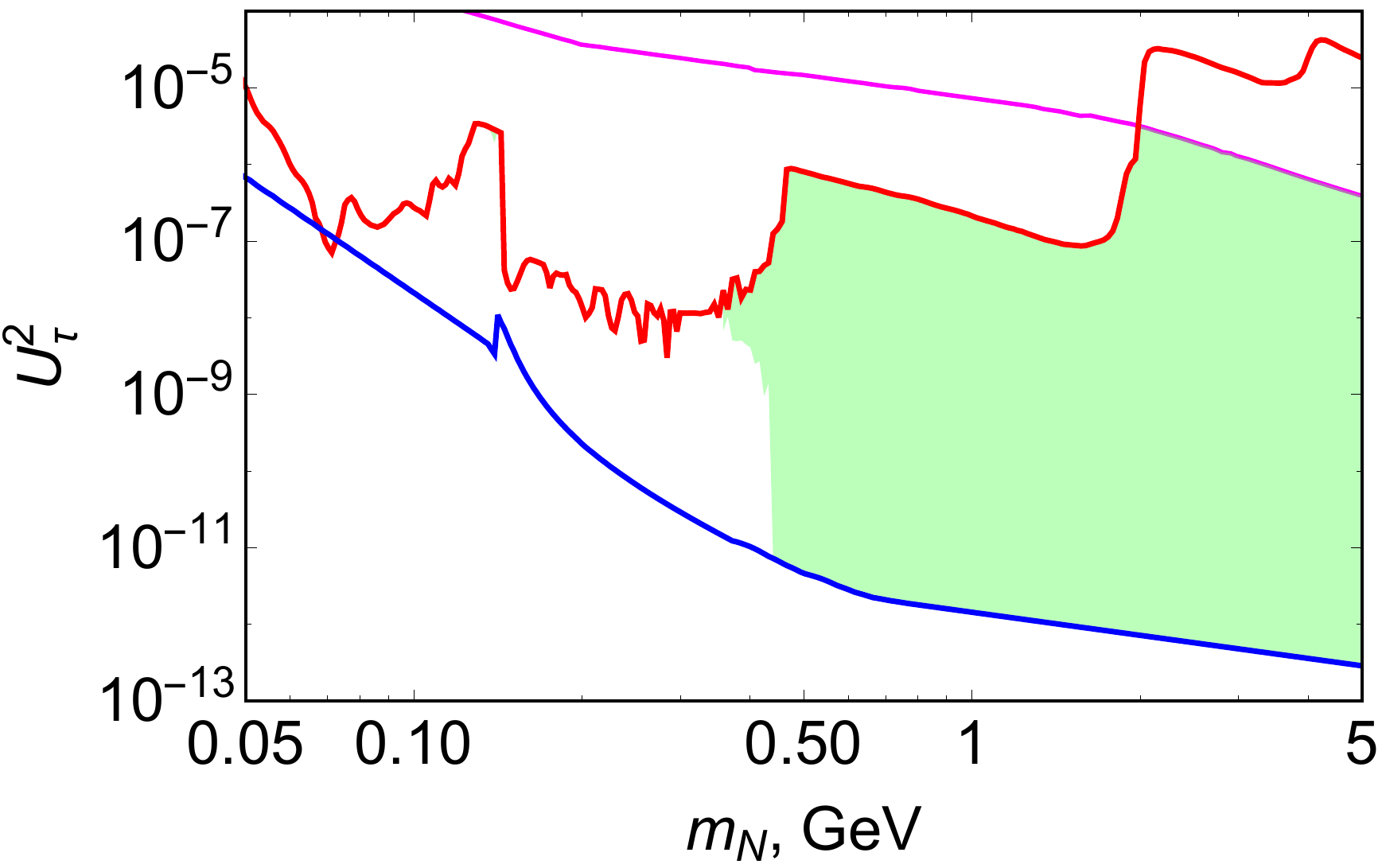}
	\caption{The parameter space of the model with two HNLs. Green points are consistent with all experimental bounds, explain neutrino data for the inverted neutrino mass hierarchy (IH) and generate the correct BAU. Other  notations are the same as in Fig.~\protect\ref{fig:NH result}.}
	\label{fig:IH result}
\end{figure}

Our results are present in Figures~\ref{fig:NH result} (for the normal hierarchy) and ~\ref{fig:IH result} (for the inverted hierarchy).
For each mass and each flavor we show a set of admissible models (green points). 
A model is selected as viable (a green point) if it explains neutrino oscillations, provides correct baryon asymmetry of the Universe and satisfies accelerator and BBN constraints, see Section~\ref{sec:procedure} for details.
The blue curve shows \emph{minimal} $U_\alpha^2$ ($U^2$) compatible with BBN in the model with 2 HNLs explaining neutrino masses and oscillations. 
The blue curve does not take into account whether other mixing angles pass selection conditions or whether BBN curve is below the accelerator curve.
The red curve shows the upper limit on $U_\alpha^2$ (correspondingly, $U^2$) compatible with accelerator searches and neutrino oscillations.

\begin{figure}[t!]
	\centering
	\includegraphics[width = 0.41\textwidth]{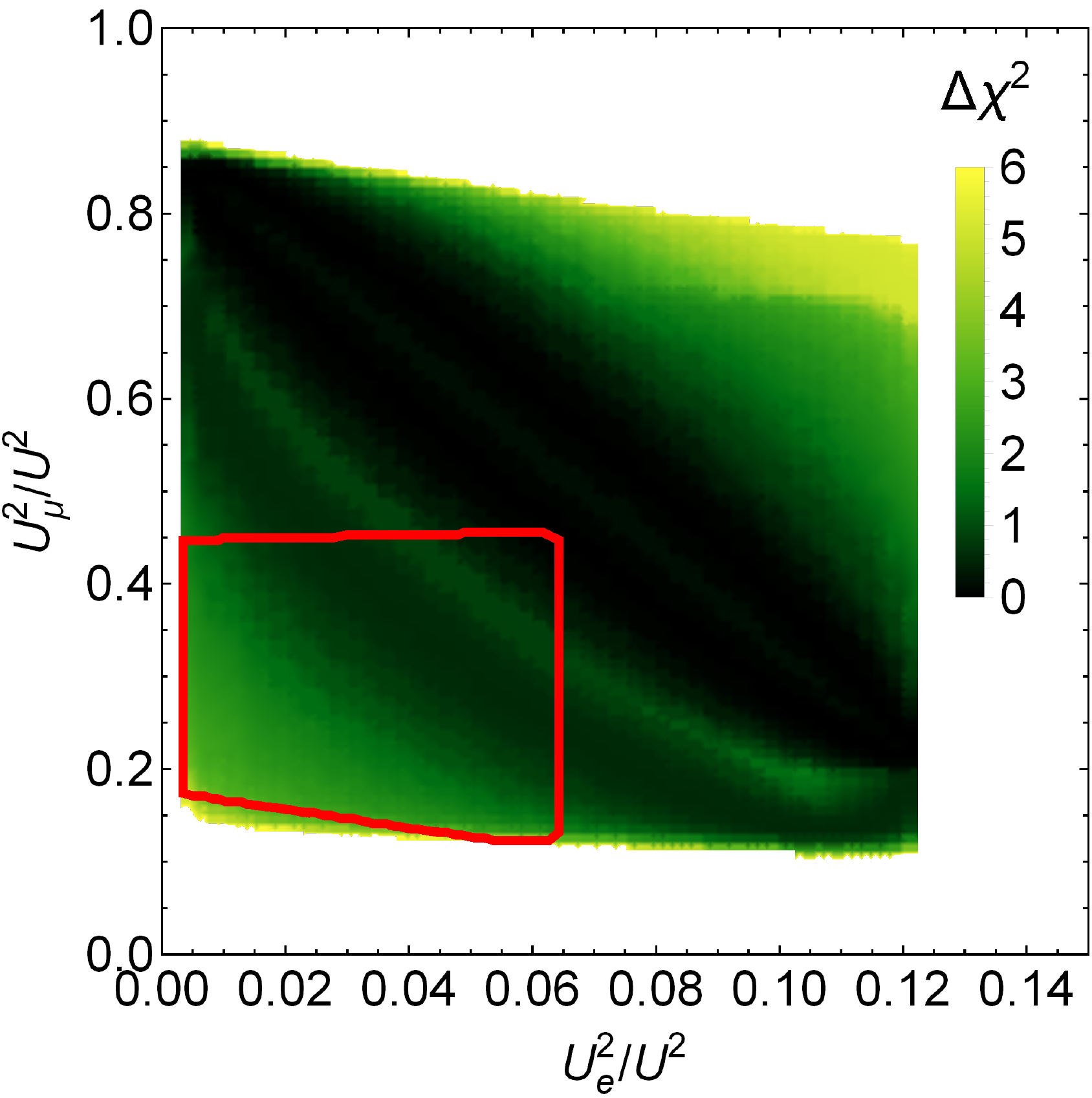}~\includegraphics[width = 0.5\textwidth]{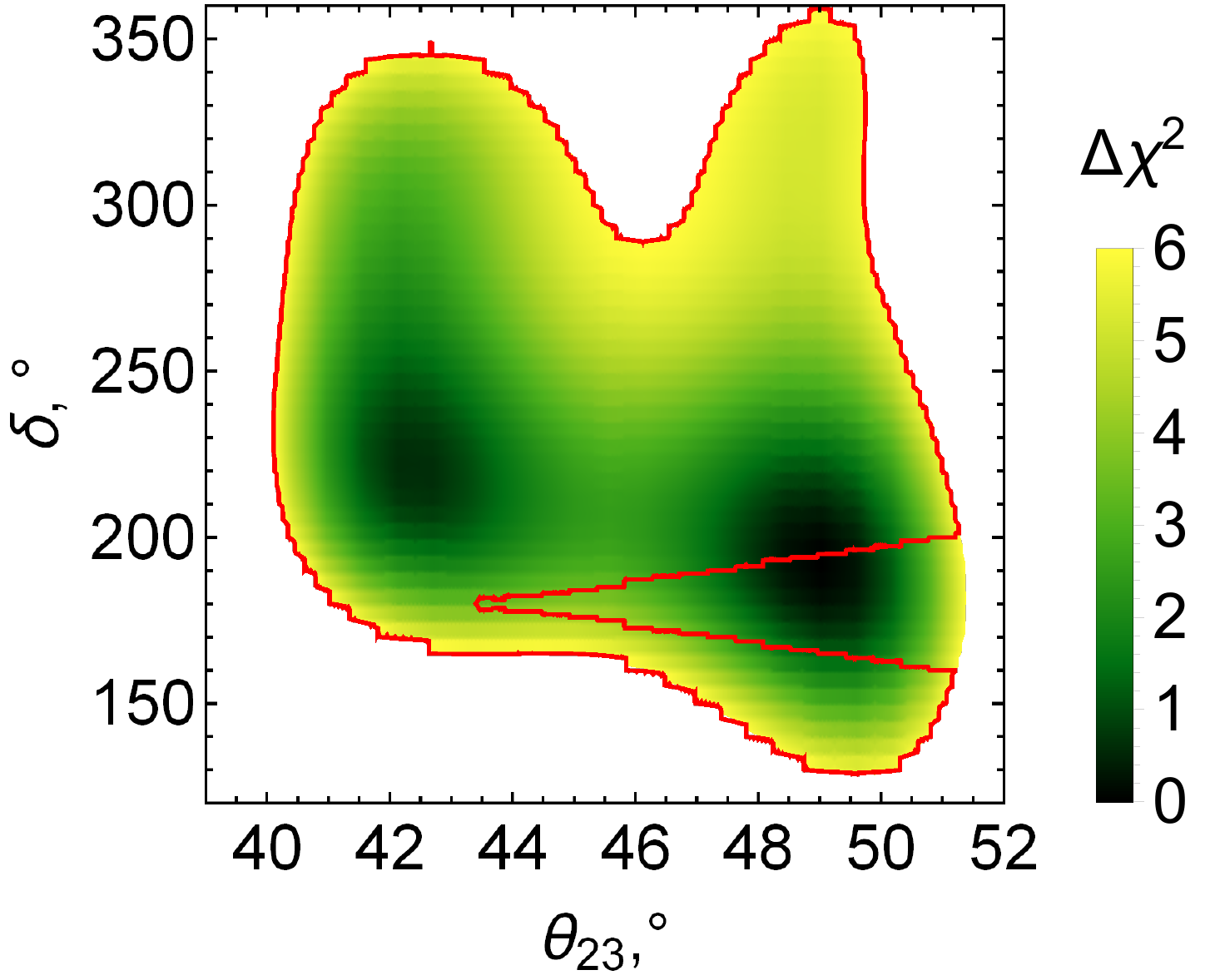}
	\caption{\textit{Left:} the allowed mixing angles in the open window $M_N \approx  140$ MeV for NH, with all of the constraints applied (red line).
	\textit{Right:} $\Delta\chi^2$ distribution in $(\delta_{\text{CP}}, \theta_{23})$ plane taken from the nuFIT 5.0~\cite{Esteban:2020cvm}. 
	The region inside the red curve corresponds to the allowed HNL models for NH and $M_N \approx 140$ MeV. 
	The CP-violating angle $\delta_{\text{CP}}$ is in degrees.
	If $\delta_{\text{CP}}$ and $\theta_{23}$ are measured to be outside the red boundary, the allowed window is excluded without a need for a dedicated search experiment.}
	\label{fig:140mev}
\end{figure}

Several comments are in order:
\begin{enumerate}
    \item Although most of the parameter space below $\approx\unit[330-360]{MeV}$ is closed, there remains an \textit{open window of viable models} for the normal hierarchy of neutrino masses (see insets in Fig.~\ref{fig:NH result}). The corresponding HNLs have masses
    \begin{equation}
        \label{eq:windwo}
	    \unit[0.12]{GeV}\le M_N \le \unit[0.14]{GeV}
    \end{equation}
    and the mixing angles (without describing the actual shape of the region)
    \begin{align*}
    \label{eq:allowedUalpha}
    3\cdot 10^{-6}\le{} &U^2 \le 15\cdot 10^{-6} 
    \\
	10^{-8}\le{} &U_e^2 \le 28\cdot 10^{-8} 
	\\
	4\cdot 10^{-7}\le{}& U_{\mu}^2 \le 26\cdot 10^{-7}
	\\
	10^{-6} \le{}& U_\tau^2 \le 12\cdot 10^{-6} 
    \end{align*}
    The existence of this window follows from the fact that the missing energy experiments do not provide strong constraints in this mass region. Experiments based on pion decays lack sensitivity due to the shrinking of available phase space, while experiments with kaon decays suffer in this region from large backgrounds, see e.g.\ recent discussion in \cite{Tastet:2020tzh}. Finally, the meson driven conversion effect which dominates the BBN constraint is not applicable in this mass range. To close the window completely one would need to improve the bounds on $U^2$ by a factor of 5-10. 
    
    The shape of the region can be seen in Fig.~\ref{fig:NH result} with the relevant experiments, while the specific (benchmark) points are listed in Table~\ref{tab:benchmarks}. The flavor ratios $U_\alpha^2/U^2$, and the PMNS parameters realized in this region of parameter space are shown in Fig.~\ref{fig:140mev}. Note that a precise determination of the PMNS parameters may be sufficient to determine the viability of this region.
    
     For the inverted hierarchy the procedure also predicts a small region at $M_N\approx \unit[0.137-0.14]{GeV}$ with $U^2$ changing slightly (we list the values for a benchmark point in Table~\ref{tab:benchmarks}). 
    \item Apart from this window, the lower mass of viable HNLs is given by
    \begin{equation}
        \begin{aligned}
        M_N>\unit[0.33]{GeV} &\quad \text{normal hierarchy}\\
        M_N>\unit[0.36]{GeV} &\quad \text{inverted hierarchy}\\
        \end{aligned}
    \end{equation}
    \begin{table}[]
        \centering
        \begin{tabular}{|c|c|c|c|c|}
        \hline
             & $M_N$ [GeV] & $U^2_e$& $U^2_\mu$ & $U^2_\tau$  \\
             \cline{1-5}
             \multirow{3}{*}{NH} 
             &0.12 &$3 \cdot 10^{-8}$ & $2\cdot 10^{-6}$ & $7\cdot 10^{-6}$
             \\
             \cline{2-5}
             & 0.14 & $2 \cdot 10^{-8}$ & $1\cdot 10^{-6}$ & $8 \cdot 10^{-6}$
             \\
             \cline{2-5}
             &$0.33$ & $8 \cdot 10^{-10}$ & $5\cdot 10^{-9}$ & $3\cdot 10^{-8}$ 
             \\
             \hline
             \multirow{2}{*}{IH} & 0.14 & $7\cdot 10^{-7}$ & $8\cdot 10^{-7}$ & $1\cdot 10^{-6}$
             \\
             \cline{2-5}
             & $0.36$ &   $2 \cdot 10^{-9}$  &$2\cdot 10^{-8}$ & $2 \cdot 10^{-8}$ \\
             \hline
        \end{tabular}
        \caption{Benchmark models for the boundary masses of the allowed regions}
        \label{tab:benchmarks}
    \end{table}
    \item For each individual flavor there are regions where accelerator bounds (red) are above the BBN limits (blue), yet the point is white. This means that such a mass is excluded by the combination of lower and upper boundaries for some other flavor.
\end{enumerate}


\subsection{Future searches}
\label{sec:future}
\begin{figure}[h!]
    \centering
    \includegraphics[width = 0.45\textwidth]{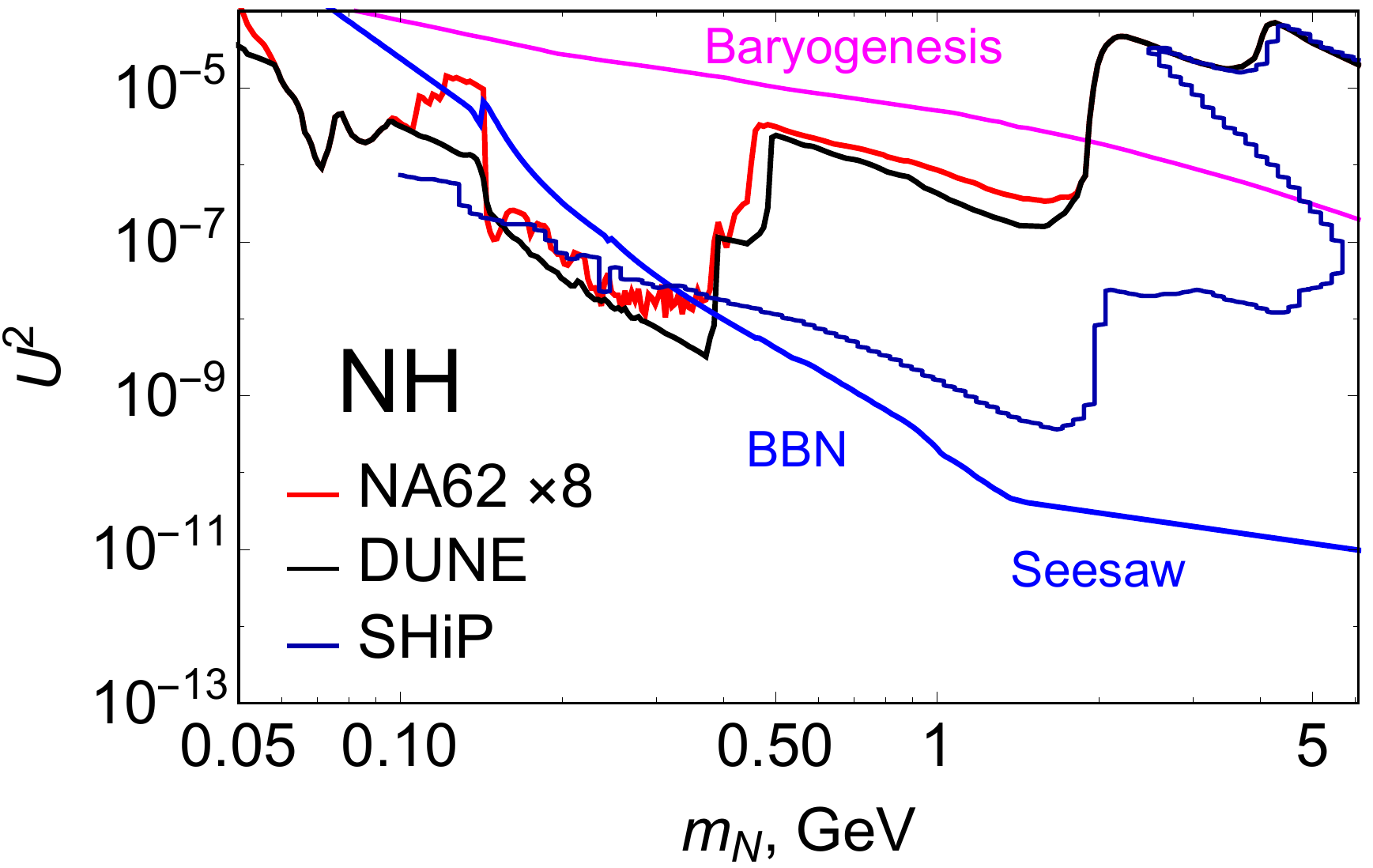}~\includegraphics[width = 0.45\textwidth]{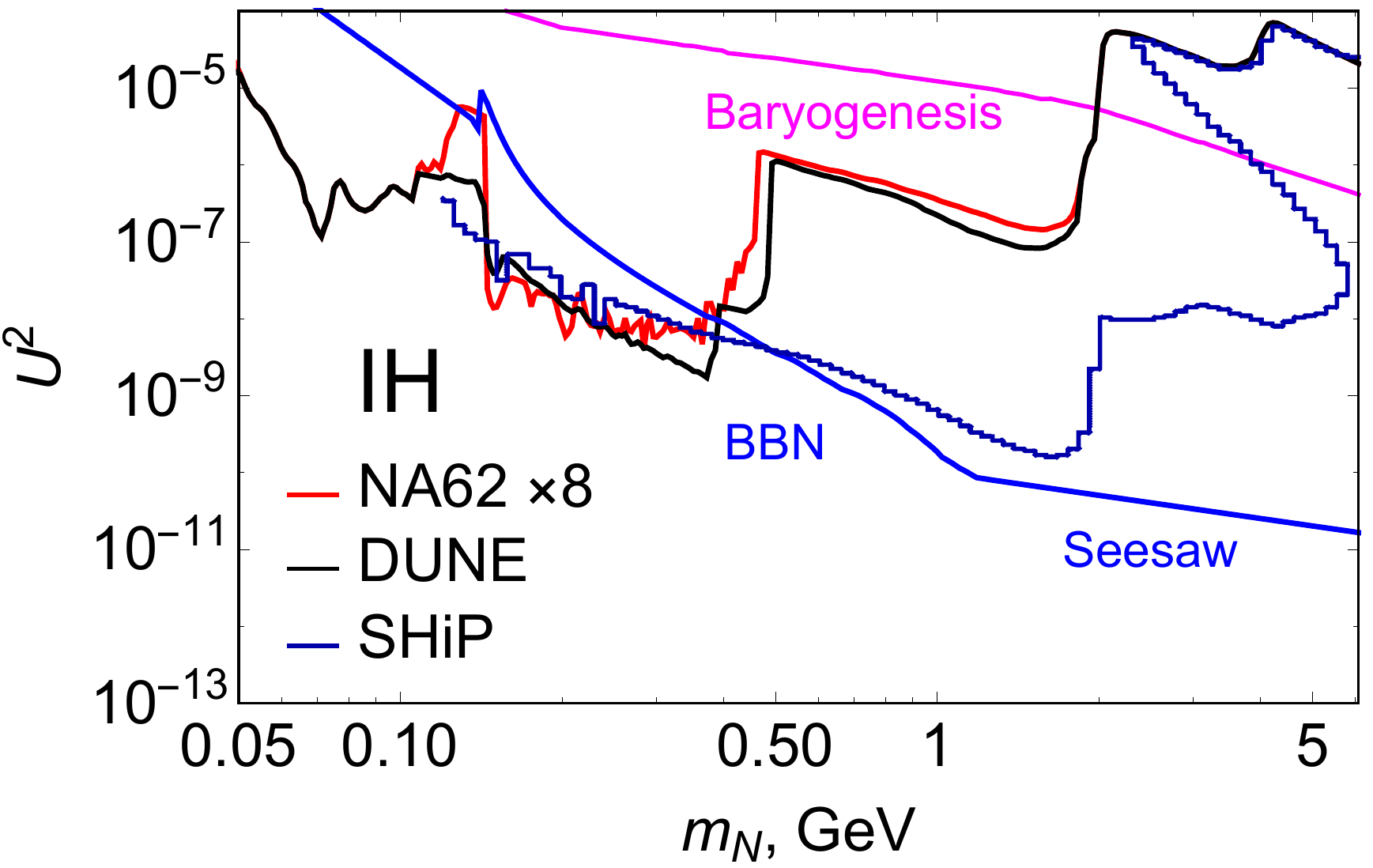}
    \caption{Parameter space of the models with two HNLs, including the projected increase of the sensitivity due to the NA62 ($\times 8$ collected data), DUNE, or SHiP experiments. The allowed points are consistent with all experimental bounds, explain neutrino data for the normal (NH) or inverted (IH)  mass hierarchy, and generate correct BAU. For NA62, the minimal mass after the pion mass window $M_N \approx m_\pi$ (for NH) becomes $M_N = \unit[0.38(0.39)]{GeV}$ for NH(IH). For DUNE, the projections are based on \cite{Coloma:2020lgy}; the minimal mass will be pushed up to $M_N \simeq \unit[0.39]{GeV}$ for both hierarchies. For SHiP, the minimal mass is $M_N \approx \unit[0.43(0.60)]{GeV}$ for NH(IH). }
    \label{fig:futuresearch}
\end{figure}

The results including the future constraints from the NA62, DUNE, and SHiP are present in Fig.~~\ref{fig:futuresearch}.

To estimate the future sensitivity of the NA62 experiment, we assume that the experiment will collect 8 times more data than has been published.\footnote{Indeed, the goal of NA62 is to collect 80 rare kaon decay $K^+\to \pi^+ \nu\bar{\nu}$ events~\cite{Ceccucci:832885}.  The existing HNL constraint~\cite{NA62:2020mcv} are based on the dataset where only $9.5$ rare kaon events are expected~\cite{NA62:2713499}. }
Assuming that both data collection and analysis strategy will not significantly change in the future and that no HNLs will be detected,  the  current limit can be scaled down as $\sqrt{8}$, taking into account that the HNL analysis is background dominated~\cite{NA62:2020mcv}.  
We see that for the normal hierarchy future NA62 measurement will not explore the HNL mass ``window'' beyond the pion mass. The remainder of the allowed parameter space is pushed to a lower mass of $M_N \gtrsim \unit[0.38(0.39)]{GeV}$ for NH(IH).

The DUNE near detector will be very sensitive to HNLs~\cite{Krasnov:2019kdc,Ballett:2019bgd,Coloma:2020lgy}. In particular, it will be able to 
push the lower bound to $M_N=\unit[0.39]{GeV}$ for both hierarchies and cover the open window at lower masses.
When estimating the sensitivity for DUNE we took $U^2_e$, $U^2_\mu$ bounds as reported in~\cite{Coloma:2020lgy} and derived $U^2$, $U_\tau^2$ bounds consistent with oscillation data.

The SHiP experiment~\cite{SHiP:2018yqc} at CERN will provide unprecedented sensitivity for heavy neutral leptons in the mass range of interest. 
Using the sensitivity matrix, provided by the SHiP collaboration \cite{SHiP:2018xqw} we have performed a full scan in the $(M_N, U^2, x_e, x_\mu)$ space to find the allowed region (determined by the number of events $n_{\text{events}}<2.3$).
The SHiP experiment will fully explore the ``window'' at low masses and push the low mass  beyond the kaon threshold: $M_N \gtrsim \unit[0.43(0.60)]{GeV}$ for NH(IH).
We note that this is a conservative estimate and the actual sensitivity will be even higher as our analysis only included HNLs coming from D-mesons \cite{SHiP:2018xqw}, while the HNLs originating from kaon decays will significantly increase the sensitivity~\cite{Gorbunov:2020rjx}.

\section{Discussion and outlook} 
\label{sec:discussion_and_outlook}

The idea that new particles need not be heavier than the electroweak scale, but rather can be light and feebly interacting draws increasing attention of both theoretical and experimental communities~\cite[see e.g.][]{Alekhin:2015byh,Beacham:2019nyx,Strategy:2019vxc}.
In particular, the idea that heavy neutral leptons are responsible for (some of the) beyond-the-Standard-Model phenomena has been actively explored in recent years, see e.g.\ \cite{Boyarsky:2009ix,Drewes:2013gca,Alekhin:2015byh,Deppisch:2015qwa} and refs.\ therein. 
This idea is motivated in the first place by the type-I seesaw model that explains neutrino oscillations. 
Furthermore, the same HNLs with nearly degenerate masses in MeV--TeV range can explain the BAU \cite[see e.g.][]{Klaric:2020lov} and refs.\ therein.

However, while theoretical developments have been focusing on the models with two or more HNLs that are mixing with different flavors, the experimental searches were concentrating on a model with a single HNL mixing with a single flavor~\cite{Liventsev:2013zz,Artamonov:2014urb,Aaij:2014aba,Khachatryan:2015gha,Aad:2015xaa,Gligorov:2017nwh,Izmaylov:2017lkv,SHiP:2018xqw,Sirunyan:2018mtv,Aad:2019kiz,NA62:2020mcv}. Such a model is simple for analysis and provides a number of useful benchmarks. Nevertheless, taken at face value it is incompatible with the observed neutrino masses and cannot generate BAU.

In this paper we address this issue. 
We recast the existing accelerator and cosmological bounds to the model with 2 HNLs with degenerate masses.  
We perform a scan over all parameter sets of the two HNL model, that  simultaneously:
\textit{(a)} explain neutrino oscillations;
\textit{(b)} are consistent with all previous non-detections at accelerators;
\textit{(c)} do not spoil predictions of Big Bang nucleosynthesis;
\textit{(d)} allow for the generation of the baryon asymmetry of the Universe.

Our main findings are as follows.
\begin{compactenum}[1.]

\item For the normal neutrino mass hierarchy, we have found an \emph{open window} for masses $0.12 - 0.14$~GeV and then for $M_N \gtrsim \unit[0.33]{GeV}$.

\item For the inverted neutrino mass hierarchy, the open window around $0.14$~GeV is very tiny with $U^2$ varying by a factor $\sim 2$ around the value $2\times 10^{-6}$. The majority of the  viable models have $M_N \gtrsim \unit[0.36]{GeV}$.

\item Future experiments,  DUNE or SHiP, will be able to fully cover the region of parameter space $0.12 - 0.14$~GeV for all values of the mixing angle. 

\item The upper mass limit above $300$~MeV will be pushed only slightly by DUNE or NA62, but will be moved beyond the kaon threshold by the SHiP experiment.

\item A precise determination of the PMNS parameters $\delta$ and $\theta_{23}$ may be sufficient for closing the  $0.12 - 0.14$~GeV window for the normal mass ordering.

\end{compactenum}

\acknowledgments{We would like to thank M.~Ovchynnikov and M.~Shaposhnikov for useful discussions. This project has received funding from the European Research Council (ERC) under the European Union's Horizon 2020 research and innovation programme (GA 694896), from the Carlsberg Foundation, and from the NWO Physics Vrij Programme “The Hidden Universe of Weakly Interacting Particles”, nr. 680.92.18.03, which is partly financed by the Dutch Research Council NWO.}

\appendix 
\section{BBN constraints on long-lived HNLs}

If HNLs possess semi-leptonic decay channels and have lifetimes $\tau_N \gtrsim \unit[0.02]{sec}$, the mesons from  HNL decays completely dominate n-p conversion rates, driving neutron-to-baryon ratio $X_n \simeq \frac 12$. The resulting abundance of Helium-4 $Y_p \simeq 2 X_n$ is then $Y_p \approx 1$, incompatible with  observations that give $Y_p < 0.2573$~\cite{Boyarsky:2020dzc}.
The HNLs with lifetimes near  the seesaw bound and masses above the pion  production threshold can have lifetimes in the range $\mathcal{O}(10^2-10^3)$~sec. 
For such  lifetimes the HNL decay products may not only affect the neutron abundance, but also destroy  already synthesized light  elements (whose production starts at Hubble times around 40~sec) -- the case that has not been analyzed in~\cite{Boyarsky:2020dzc}.

Below we demonstrate that for all values of HNL masses/lifetimes compatible with neutrino oscillations, such HNLs lead to an overproduction of Helium-4 or other light elements and therefore the region $\tau_N \gtrsim \unit[40]{sec}$ and $M_N > m_\pi$ is also excluded from BBN.
The details of the analysis will be presented elsewhere~\cite{very_long_HNLs}.

Indeed, all the neutrons in the primordial plasma will either decay or bind into light elements (deuterium, Helium-3, Helium-4, etc).
The presence of pions in the plasma effectively ``prevents'' neutrons from decaying because the rate of $n + \pi^+ \to p + \pi^0$ exceeds both the Hubble expansion rate and the decay rate  $n\to p + e^- + \bar\nu_e$.
As a result, decays (and other weak processes) can be ignored until Hubble times $\sim 10^6$~sec, leading to the following equation of neutron balance: 
\begin{equation}
    \label{eq:neutron_balance}
    X_n^{\rm (free)} + X_D +  X_{\isotope[3]{He}} + 2X_{\isotope[4]{He}}+\dots\approx \frac 12
\end{equation}
The cross-sections of all reactions that change abundances in~\eqref{eq:neutron_balance} ($n\leftrightarrow p$ conversion by pions, nucleosynthesis, dissociation of light nuclei by pions, etc) are  of the same order. 
Therefore, the rates of various reactions are determined solely by the concentrations. Without going into details (see~~\cite{very_long_HNLs}) there are two qualitative regimes: if 
 the instantaneous concentration of pions is $n_\pi \gtrsim n_B$ -- the pions will efficiently destroy the synthesized nuclei and all terms in the l.h.s.\ of Eq.~\eqref{eq:neutron_balance} will end up being of the same order $X_n^{\rm (free)} \sim X_D \sim  X_{\isotope[3]{He}} \sim X_{\isotope[4]{He}}\sim \mathcal{O}(1)$.
 If, on the other hand, the instantaneous concentration of pions is small, $n_\pi < n_B$, -- most of the neutrons will bind into the nuclei, leading to $X_{\isotope[4]{He}}\sim 1$ and $X_n^{\rm (free)} \ll 1$.
 Both cases are  incompatible with experimentally observed  abundances $X_D \sim X_{\isotope[3]{He}}\sim 10^{-5}$ and $X_{\isotope[4]{He}} \sim 0.0643$.\footnote{We remind that the mass fraction $Y_p = 4 X_{\isotope[4]{He}} = 2X_n$.} 

Finally, few words should be said about long-lived HNLs with $M_N < m_\pi$.
The influence of such particles on BBN has been analyzed in a number of recent works~\cite[see e.g.][]{Ruchayskiy:2012si,Sabti:2020yrt}, providing an upper bound on the HNL lifetime that is below the seesaw limit.
Near the seesaw boundary the HNLs are long-lived, so they can survive till the onset of nuclear reactions and their decay products can dissociate light nuclei.
The recent analysis of~\cite{Domcke:2020ety} based on \cite{Hufnagel:2018bjp} (see also \cite{Forestell:2018txr}) demonstrated that MeV mass HNLs with lifetimes exceeding the seesaw bound are \emph{excluded} from cosmological observations (BBN plus CMB) and therefore no ``open window'' exists below the seesaw line but above the limits of \cite{Ruchayskiy:2012si,Sabti:2020yrt}.

\bibliographystyle{JHEP}
\bibliography{main}

\end{document}